\documentclass[10pt]{iopart}

\usepackage{iopams}
\usepackage{setstack}
\usepackage{graphicx}
\usepackage{float}
\usepackage{xcolor}
\usepackage{hyperref}

\newcommand{\avg}[2]{\ensuremath{\left\langle #1 \right\rangle_{#2}}}
\newcommand{\dst}[0]{\ensuremath{\mathrm{d}}}
\newcommand{\pdev}[3]{\ensuremath{ \frac{ \partial^{#3} #1 }{ {\partial #2}^{#3} } } }
\newcommand{\arcosh}[0]{\ensuremath{ \mathrm{arccosh} }}
\newcommand{\tobs}[0]{\ensuremath{T}}

\newcommand{\revjg}[1]{\textcolor{black}{#1}}

\definecolor{mygreen}{rgb}{0.0,0.6,0.45}
\newcommand{\rlj}[1]{{\color{black}#1}}

\begin{document}

\title[Dynamical phase transitions for the activity biased Ising model]
{Dynamical phase transitions for the activity biased Ising model in a magnetic field}
\author{Jules Guioth$^1$, Robert L. Jack$^{1, 2}$}
\address{$^1$ Department of Applied Mathematics and Theoretical Physics, University  of Cambridge, Wilberforce Road, Cambridge CB3 0WA, United Kingdom}
\address{$^2$ Department of Chemistry, University of Cambridge, Lensfield Road, Cambridge CB2 1EW, United Kingdom}
  
\ead{jules.guioth@damtp.cam.ac.uk}

\begin{abstract}
 We consider large deviations of the \revjg{dynamical} activity \revjg{-- defined as the total number of configuration changes within a time interval --} for mean-field and one-dimensional Ising models, in the presence of a magnetic field.
 \rlj{We identify several dynamical phase transitions that appear as singularities in the scaled cumulant generating function of the activity.  In particular,}
 we find low-activity ferromagnetic states and a novel high-activity phase, with associated first- and second-order phase transitions.
 The high-activity phase has a negative susceptibility to the magnetic field.
 In the mean-field case, 
 we analyse the dynamical phase coexistence that occurs on first-order transition lines, including the optimal-control forces that reproduce the relevant large deviations.
 In the one-dimensional model, we use exact diagonalisation and cloning methods to perform finite-size scaling of the first-order phase transition at non-zero magnetic field.
\end{abstract}


\maketitle


\section{Introduction}

Understanding dynamical fluctuation phenomena is important in many physical contexts.  Building on insights from fluctuation theorems~\cite{PhysRevLett.74.2694,Jarzynski1997}, tools of large-deviation theory~\cite{denH-book,derrida2007non,touchette2009large,Bertini2015} are now commonly used to gain insight into non-equilibrium fluctuations.  In particular, large deviations of time-integrated quantities have links to ergodic behavior and are intrinsically linked with the emergence of long time scales, as reviewed in~\cite{Jack2019}.  
Such methods have been applied to current fluctuations in driven systems \cite{bodineau2004current,derrida2007non,Gingrich2016}, and to the slow relaxation of glassy materials~\cite{garrahan2007dynamical,garrahan2009first,hedges2009dynamic,pinch17,Malins2012-sens}.

A striking aspect of this theory is the existence of dynamical phase transitions~\cite{Bodineau2005,lecomte2007thermodynamic,garrahan2007dynamical,baek2017dynamical}
whose physical signature is a qualitative change in the mechanism for large fluctuations, often accompanied by spontaneous symmetry breaking.
Within the theory, these transitions correspond to singularities in certain large-deviation functions, which are analogous to the free energy or entropy in equilibrium statistical mechanics.

This article revisits a prototypical model system where such phase transitions occur -- the Ising model with Glauber dynamics~\cite{jack2010large,loscar2011thermodynamics,garrahan2009first,lecomte2007thermodynamic,vasiloiu2019trajectory}.  We consider the mean-field version of the model (similar to~\cite{lecomte2007thermodynamic}) and the model in one dimension (similar to~\cite{jack2010large}).  The phase transitions that we consider are related to time-integrated measurements of dynamical activity, \rlj{defined} as the number of times that the system changes its state over a long time interval $[0,T]$.  
\rlj{This extends previous work in several ways, the most notable being the existence of new phase transitions (both} first-order and second-order) that occur for systems in magnetic fields, when considering large fluctuations with high activity.

\rlj{Since we consider the Ising model at equilibrium, the system is time-reversal symmetric.  The activity is also a time-reversal symmetric quantity, it is related to the \emph{frenesy} \cite{maesFrenesy2020} which is the time-reversal symmetric part of the dynamical action.\footnote{%
\rlj{The activity considered here is that of~\cite{garrahan2007dynamical,garrahan2009first} which is different from the activity defined in~\cite{maesFrenesy2020}.  Nevertheless, these quantities are correlated with each other and have similar physical content, which is to quantify how much motion is taking place in a given trajectory, see also~\cite{jack2006,Fullerton2013}.}%
}
Hence, the large deviations considered in this work occur by mechanisms that are time-reversal symmetric.  This may be contrasted with the entropy production, which is the time-reversal antisymmetric part of the action; its large-deviation behaviour obeys fluctuation theorems~\cite{PhysRevLett.74.2694,Jarzynski1997,lebowitz1999gallavotti,Crooks2000}, which can be used to quantify the difference in probability between a fluctuation mechanism and its time-reversed counterpart.    

Large deviations of the dynamical activity (and other time-reversal symmetric quantities) have been analysed extensively in models of glasses~\cite{garrahan2007dynamical,garrahan2009first,hedges2009dynamic,Speck2012-sens,Fullerton2013,Jack2014east,Turner2015}.
These studies observed first-order dynamical phase transitions, when considering large deviations where the activity is lower than its typical value. The low-activity phase consists of low-energy glassy configurations which come from metastable states with long (but finite) lifetimes~\cite{Jack2011prep,Malins2012-sens}.   Dynamical phase transitions have also been demonstrated in Ising models ~\cite{garrahan2009first,jack2010large,loscar2011thermodynamics}, leading to long-ranged ferromagnetic order, even in one-dimension.  In addition, large deviations of the activity are linked to dynamical phase transitions in exclusion processes~\cite{appert2008,Lecomte2012}, which may be either second-order or first-order~\cite{baek2017dynamical}, and are related to slow hydrodynamic modes. 

The behaviour in glassy models has links with that of Ising-like models.  Qualitatively, one finds that low-activity dynamical phases are characterised by  enhanced order -- which is ferromagnetic in the Ising model but has a more complex form in glasses~\cite{Malins2012-sens,Fullerton2014,BerthierJack2015}.  Also, in cases where dynamical critical points have been characterised in glassy models, they are in the Ising universality class~\cite{Elmatad2010,Elmatad2013}.  This is consistent with general theoretical arguments: the activity is a scalar order parameter so one may expect the critical behaviour to be described by a field theory of $\phi^4$ type, defined in $(d+1)$-dimensional space-time, leading to Ising-like critical behaviour.  On the other hand, for equilibrium phase transitions, the fact that glassy materials are disordered can result in critical points with characteristics of random-field Ising models~\cite{franz2013glassy,biroli2014random}, indicating that caution is required before assuming that their dynamical phase transitions are Ising-like in all cases.

The present article investigates large deviations of the activity in Ising models, including regimes that were not studied before.  The aim is to understand better what kinds of dynamical phase transition can occur, what causes them, and what is their physical interpretation.  Such results are valuable as theoretical context, which can be compared with existing results for glassy systems.  We find critical points that occur in magnetic fields, leading to dynamical phases that are not related by any symmetry of the model -- this is more similar to the glassy case than the zero-field critical point of the Ising model.  We also find dynamical phase transitions at high activities, distinct from most cases studied so far.
These phases also have magnetisations that are anti-parallel to the applied magnetic field, which is reminiscent of spherical models~\cite{van2010second}, in which these phases were identified as \emph{anomalous} (because of their negative magnetic susceptibility).
}%

The mean-field model considered here leads to a simpler analysis than the spherical model of~\cite{van2010second}, allowing a clearer characterisation of these phases.  We also demonstrate numerically that a similar phase transition also occurs in the one-dimensional model, using a combination of exact diagonalisation and cloning methods~\cite{giardina2006direct,lecomte2007numerical}.  We use these transitions to discuss behaviour at (and close to) dynamical phase coexistence, showing that some results for specific systems~\cite{nemoto2014finite,nemoto2017finite,JackNemoto2019} can also be generalised to this case.

\rlj{%
Taken together, our results provide further examples of the rich phenomenology associated with large deviations of the activity, even in simple models.  Setting aside the details, one question that remains is: What physical insight is available from studying these rare events, and the associated dynamical phases?  In contrast to correlation functions involving the frenesy and activity (for example the covariances that appear in non-equilibrium linear response theories~\cite{maesFrenesy2020}), large-deviation properties cannot be measured directly, except by waiting for rare events to occur and assembling the associated histograms~\cite{pinch17}, or by measuring high-order cumulants~\cite{flindt2013trajectory}.   However, qualitative features of dynamical phases can be useful for understanding metastable states and other slow processes.
For example, in the glassy context, the  low-activity phase is characterised by long-ranged order in space and time~\cite{hedges2009dynamic,Elmatad2010}, which can be interpreted as a long-ranged analogue of the (short-ranged) dynamical heterogeneities that are characteristic of supercooled liquids~\cite{Chandler2010}.   The subtle structural order of the low-activity phase also lends insight into their metastability~\cite{Malins2012-sens,Fullerton2014,BerthierJack2015}.   

For the high-activity phases considered here, we again find structural order, which we attribute to a competition between two effects.  On the one hand, individual spins should change their state frequently (high activity); on the other hand, large-deviation events happen by the \emph{least unlikely} of all possible mechanisms, which is often associated with states from which relaxation to equilibrium is slow~\cite{Jack2010rom}.  One mechanism that is consistent with both effects is to localise the system near a saddle point of the free energy, leading to a small free-energy gradient (hence slow global relaxation) but maintaining high local activity.  Our results are generally consistent with this idea, details are given below.
As a general conclusion, our study supports the view that while properties of dynamical phases may be hard to anticipate (for example, magnetisation opposite to an applied field), the origins of such effects can be traced back to systems' physical properties, particularly their free energy landscapes and the presence of slow physical processes.
Additionally, there are connections between large-deviation theory and ideas of optimal control~\cite{chetrite2015variational,Jack2019}, which relate large-deviation behavior to (non-linear) responses, for a particular set of applied (control) forces.
 Our hope is that by improving our understanding of these links in simple models, we become better equipped to interpret the large-deviation behaviour of more complex systems like glasses~\cite{hedges2009dynamic} and active matter~\cite{nemoto2019optimizing}.%
}

This article is organised as follows: We introduce in \sref{sec:theory} the main theoretical ingredients of our analysis. 
In \sref{sec:mfim}, we discuss analytically the Mean-Field (Curie-Weiss) version of the Ising model. 

Section \ref{sec:Ising1d} discusses the one dimensional Ising model and shows that the latter displays a similar phase diagram as the Mean-Field version.
Finally, we summarise our conclusions in \sref{sec:conclusion}.  Some technical results are presented in Appendices.

\section{Theory: activity biased dynamics of the Ising model}\label{sec:theory}

This section explains how standard methods of large-deviation theory are applied to the Ising model.  
Further detail and context for the methods can be found in~\cite{chetrite2015nonequilibrium,chetrite2015variational,Jack2019}.

\subsection{Model}
\label{sec:model}

We consider Ising models where the $i$th spin is $\sigma_i=\pm1$.  There are $N$ spins in total and the overall configuration of the system is $\bsigma=(\sigma_i)_{i=1}^N$.  For the one-dimensional variant of the model, the energy is
\begin{equation}
  \label{eq:def:energy_ising}
  E(\bsigma) =  \sum_{i=1}^N \left( - J \sigma_{i} \sigma_{i+1} - h \sigma_{i} \right)
\end{equation}
where $J$ is the coupling constant and $h$ the magnetic field.  We take periodic boundary conditions so it is understood that $\sigma_{N+1}=\sigma_1$.  
The magnetisation of the system is
\begin{equation}
m(\bsigma) = \frac{1}{N} \sum_{i=1}^N \sigma_i \; .
\end{equation}
We also consider a mean-field variant of the model for which the energy is
\begin{equation}
  \label{eq:def:energy_mf}
  E(\bsigma) = N {\cal E}( m(\bsigma) ), \qquad {\cal E}(m) = -Jm^2 - hm   \; .
\end{equation}
The inverse temperature is $\beta$ and the associated Boltzmann distribution is 
\begin{equation}
P_{\mathrm{eq}}(\bsigma) = z^{-1}\exp\left(-\beta E(\bsigma)\right) \; ,
\end{equation}
where $z$ is the (thermodynamic) partition function.

We consider Markov jump dynamics in continuous time.  The jump rates respect detailed balance with respect to $P_{\rm eq}$, so the transition rate from state $\bsigma$ to $\bsigma'$ takes the form
\begin{equation}
  \label{eq:def:general_transition_rates}
  w(\bsigma'| \bsigma) = a(\bsigma, \bsigma') \exp\left\{ - \frac{\beta}{2} \left[ E(\bsigma') - E(\bsigma) \right] \,  \right\}
\end{equation}
where the function $a$ is symmetric, that is $a(\bsigma, \bsigma')=a(\bsigma', \bsigma)$.  It may be interpreted as a mobility \cite{kaiser2018canonical,maes2007and}.  We focus here on Glauber dynamics, which corresponds to
\begin{equation}
  \label{eq:def:glauber_type_transition_rates}
  a(\bsigma, \bsigma') = \frac{1}{\cosh\left[ \frac{\beta}{2} \left( E(\bsigma') - E(\bsigma) \right) \right]} \; .
\end{equation}
Define also the escape rate from state $\bsigma$ as
\begin{equation}
r(\bsigma) = \sum_{\bsigma' (\neq \bsigma)}w(\bsigma'|\bsigma) \; .
\label{eq:esc-rate}
\end{equation}
Given a system in state $\bsigma$, the time until the next spin flip is exponentially distributed with mean $r(\bsigma)^{-1}$.

A trajectory of the system on the time interval $[0,T]$ is denoted by
$\Theta_{T} = \{ \bsigma(t) \}_{t \in [0, \tobs]}$.  Let ${\cal K}[\Theta_T]$ denote the number of jumps (spin flips) in this trajectory, and the empirical (average) jump rate is
\begin{equation}
\label{eq:def:activity}
\mathcal{N}[\Theta_{\tobs}] = \frac{1}{T} {\cal K}[\Theta_T] \; .
\end{equation}
(We emphasise that $T$ denotes the duration of the dynamical trajectory, there should be no confusion with the temperature of the system, which is $1/\beta$.)
The empirical jump rate reflects the amount of dynamical activity in a trajectory.  An alternative characterisation of the activity is given by the time-averaged escape rate
\begin{equation}
  \label{eq:def:time-avg_esc_rate}
  \mathcal{R}[\Theta_{\tobs}] = \frac{1}{\tobs}\int_{0}^{\tobs} r(\bsigma(t)) \,  \mathrm{d}t \, .
\end{equation}
Note that these observables are of different types, in that ${\cal R}$ is an integral of a one-time quantity, while ${\cal N}$ depends on jumps between states.  However, the statistical properties of these observables are intimately connected, because of the underlying Poisson dynamics of the jumps (see \cite[App. B]{garrahan2009first}).

\subsection{Conditioned and biased ensembles: large-deviation analysis}

Let ${\cal A}$ denote a generic measure of dynamical activity, for example ${\cal N}$ or ${\cal R}$ as defined above.
We consider the physical behaviour of a system, under the condition that ${\cal A}[\Theta_T]$ takes a non-typical value, for large $T$.
The theory of large deviations can then be used to analyse the behaviour $T\to\infty$, see for example~\cite{lecomte2007thermodynamic,garrahan2009first,touchette2009large,chetrite2015variational}.  We briefly summarise the relevant theory.

Since the system is a finite Markov chain, the activity obeys a large-deviation principle, namely
\begin{equation}
  \label{eq:LDP_time-integrated_activity}
  P(\mathcal{A}[\Theta_{\tobs}] \approx a) \underset{\tobs\to\infty}{\sim} e^{-\tobs I(a)} \, ,
\end{equation}
where $I(a)$ is the large-deviation function (or rate function).  Denote the typical value of ${\cal A}$ by $a^\ast$.  
At equilibrium, most of the observed trajectories have $\mathcal{A}\approx a^{\ast}$, 
so that
$I(a^{\ast}) = 0$ and $I'(a^{\ast})=0$. Other values of $a$ involve large fluctuations whose probabilities are quantified by $I(a)$. 

In addition to the probability of such events, it is also possible to characterise their mechanism -- that is, the behaviour of the (very unlikely) trajectories $\Theta_T$ that realise the non-typical activity $a$.  To this end, define a conditional probability distribution for trajectories as
\begin{equation}
  \label{eq:cond_ensemble}
  P(\Theta_{\tobs} | \mathcal{A}[\Theta_{\tobs}] = a) \propto P(\Theta_{\tobs}) \delta(\mathcal{A}[\Theta_{\tobs}] - a) \; .
\end{equation}
where the constant of proportionality is fixed by normalisation.
In practice, such conditioned distributions may not be convenient to handle, so one introduces a corresponding \emph{biased} ensemble --- sometimes called the $s$-ensemble --- defined as
\begin{equation}
  \label{eq:biased_ensemble}
  P_{s}(\Theta_{\tobs}) = \frac{P(\Theta_{\tobs})}{Z(s,\tobs)} e^{-s \tobs \mathcal{A}[\Theta_{\tobs}]} \; ,
\end{equation}
where the normalisation constant
\begin{equation}\label{eq:partition_function_biased_ens}
  Z(s,\tobs) = \left\langle \exp(-s \tobs \mathcal{A}[\Theta_{\tobs}]) \right\rangle  
\end{equation}
is similar to the partition function in equilibrium statistical mechanics.  The distributions (\ref{eq:cond_ensemble},\ref{eq:biased_ensemble}),
are related to each other, just as microcanonical and canonical ensembles are related in thermodynamics~\cite{garrahan2009first,chetrite2015nonequilibrium}.  We focus here on the \emph{biased} ensemble.
Note that 
$Z(s,\tobs)$ is the moment generating function for ${\cal A}[\Theta_T]$.  It behaves for large $T$ as
\begin{equation}  
  \label{eq:partition_function_Psi}
    Z(s,\tobs) \underset{\tobs \to \infty}{\sim} e^{\tobs\Psi(s)}    
\end{equation}
where $\Psi(s)$ is analogous to the (negative of the) free-energy in the canonical ensemble.  This quantity also depends (implicitly) on the system size $N$, we sometimes make this explicit by writing $\Psi_N$.
We will be interested below in dynamical phase transitions that appear in the limit where $N$ and $T$ are both very large.  We therefore define
\begin{equation}
\psi(s) = \lim_{N\to\infty} \lim_{T\to\infty} \frac{1}{NT} \log Z(s,T)  \; .
\label{eq:small-psi}
\end{equation}
The function $\Psi_N$ is guaranteed to be analytic (because $N$ is finite) but $\psi$ may have singularities, which correspond to dynamical phase transitions.  
As discussed in~\cite{Jack2019,JackNemoto2019}, one expects quite generally that the two limits commute in (\ref{eq:small-psi}), but other properties of the biased ensemble can depend strongly on the relative size of $N$ and $T$.
\subsection{Dynamical free energy and optimally controlled process}
\label{sec:theory_markov_jump}

To analyse the biased ensemble, we define an operator (or matrix) whose largest eigenvalue coincides with $\Psi(s)$.   This matrix is denoted by ${\cal W}_s$.  
For the case where the dynamical activity ${\cal A}={\cal N}$ (the number of spin flips), the matrix elements of ${\cal W}_s$ are 
\begin{equation}
  \label{eq:biased_operator}
  (\mathcal{W}_{s})_{\bsigma', \bsigma} = e^{-s}w(\bsigma'|\bsigma) - r(\bsigma)\delta_{\bsigma', \bsigma} \, .
\end{equation}
where $\bsigma,\bsigma'$ are configurations of the model. (The size of the matrix is $2^N\times 2^N$, where $N$ is the number of spins.  It can be interpreted as an operator that acts on vectors $p$ with elements $p_{\bsigma}$.  Then $p$ corresponds to an (unnormalised) probability distribution.)  

The largest eigenvalue of $\mathcal{W}_{s}$ can be alternatively characterised by a variational formula, which is also related to optimal control theory.  To this end, define a new Markov jump process (\emph{controlled process}) where the transition rates $w$ are modified as
\begin{equation}\label{eq:transition_rates_opt_process}
  w^{\rm con}(\bsigma' | \bsigma) = a(\bsigma, \bsigma') \sqrt{\frac{\mu(\bsigma')}{\mu(\bsigma)}} \, ,
\end{equation}
where $\mu$ is a function that assigns a positive number to each state $\bsigma$.   
It is useful to normalize these numbers as $\sum_{\bsigma} \mu(\bsigma)=1$.  
Then $\mu$ is a probability distributions over the configurations of the model, and the controlled transition rates (\ref{eq:transition_rates_opt_process}) respect detailed balance with respect to this distribution. {Then one has \cite[Eq. (27)]{garrahan2009first}}
\begin{equation}
\Psi(s)  =  \underset{\mu}{\max}\left[ e^{-s} \avg{r^{\rm con}}{\mu} - \avg{r}{\mu}  \right]  \; .
\label{eq:psi-var}
\end{equation}
Here, $\langle f \rangle_\mu = \sum_{\bsigma} f(\bsigma) \mu(\bsigma)$ indicates the average of the $\bsigma$-dependent observable $f$ with respect to the distribution $\mu$; the escape rate $r$ is given by (\ref{eq:esc-rate}), and similarly 
\begin{equation}\label{eq:escape_rate_opt_process}
  r^{\rm con}(\bsigma) = \sum_{\bsigma' (\neq \bsigma)} w^{\rm con}(\bsigma' | \bsigma )
\end{equation}
is the escape rate for the controlled process.

Let $\mu^\ast$ be the distribution that achieves the maximum in (\ref{eq:psi-var}).  Then
the right eigenvector of ${\cal W}_s$ has elements $[\mu^*(\bsigma)P_{\rm eq}(\bsigma)]^{1/2}$ and its left eigenvector has elements 
$[\mu^*(\bsigma)/P_{\rm eq}(\bsigma)]^{1/2}$.  Using this $\mu^\ast$ in (\ref{eq:transition_rates_opt_process}) yields transition rates for the \emph{optimally controlled process}.  This is a Markov jump process that generates trajectories from a distribution that is very close to the biased ensemble (\ref{eq:biased_ensemble}), see~\cite{jack2010large,chetrite2015nonequilibrium,chetrite2015variational}.  As such, it captures the mechanism of large-deviation events with non-typical values of ${\cal N}[\Theta_T]$.

\rlj{In this sense, the large-deviation events that we consider can also be interpreted as (nonlinear) responses to the optimal control forces.}
\revjg{The required modification to the natural dynamics follows from (\ref{eq:transition_rates_opt_process}) as
    \begin{equation}
      \label{eq:optimal_control_forces_explicit}
      w^{\rm con}(\bsigma'|\bsigma) = w(\bsigma'|\bsigma)e^{-(\beta/2)[U^{\rm con}(\bsigma') - U^{\rm con}(\bsigma)]} \, ,
    \end{equation}
    where
    \begin{equation}
      \label{eq:optimal_control_forces_def}
      U^{\rm con}(\bsigma)= \frac{-1}{\beta} \log \frac{\mu^{*}(\bsigma)}{P_{\rm eq}(\bsigma)} \; .
    \end{equation}
    is the (optimal) control potential, whose gradients are the control forces.
We emphasise that these optimal forces are not typically realisable in experiments and may not correspond to physically natural perturbations.
However, the controlled system is Markovian; in the cases considered here it is also time-reversal symmetric.}
 
\revjg{
  As will become clearer in the detailed study of the Mean-Field model, the dynamical free energy \eref{eq:psi-var} is determined by a variational principle that involves two contributions. For $s\gg 1$, one expects that $\Psi(s)$ is determined by states that minimise the average escape rate $\avg{r}{\mu}$. On the other hand, when $s\ll -1$, $\Psi(s)$ is mostly determined by states that maximise the escape rate for the controlled process $\avg{r^{\rm con}}{\mu}$.
  Therefore, one may adopt the analogy with the equilibrium statistical mechanics and interpret the escape rate $r$ as an \emph{energy}, the escape rate of the controlled system $r^{\rm con}$ as an \emph{entropy} and $e^{-s}$ as an effective temperature controlling the balance between both terms.
  }


\section{Mean-Field Ising Model: analytical study}\label{sec:mfim}

We consider the mean-field variant of the Ising model, whose energy function was given in (\ref{eq:def:energy_mf}).  
\rlj{We set the inverse temperature $\beta=1$ throughout this Section, without any loss of generality.}
Flipping spin $i$ involves an energy change $\Delta E = 2\sigma_{i}(2Jm + h) - 4J/N$, where $\sigma_{i}$ is the state of the spin just before the flip. As expected for a Mean-Field model, all the spins behave as if they were independent entities interacting with an effective external field $2Jm + h$. Hence, the probability to flip a spin depends only on its own value and the dynamical evolution can be simplified into a Markov chain for the total magnetisation. It is convenient to introduce a function $\gamma$ to encapsulate effects of the microscopic mobility $a$.  Then the transition rates for the magnetisation are
\begin{equation}\label{eq:transition_rates_mfim}
  \fl
  w_{N}(m' | m ) =
  \cases{
    N \gamma\Big( 2Jm + h + \case{2J}{N} \Big) \frac{1 - m}{2}  {\rm e}^{  \left[ 2Jm + h + \case{2J}{N} \right] } & \text{if} $ m'=m+2/N $  \\    
    N  \gamma\Big( 2Jm + h - \case{2J}{N} \Big) \frac{1 + m}{2} {\rm e}^{  \left[-2Jm - h + \case{2J}{N} \right] } & \text{if} $ m'=m-2/N $
}
  \, , 
\end{equation}
[for other values of $m,m'$ then $w_N(m|m')=0$.  If the microscopic dynamics are Glauber as in (\ref{eq:def:glauber_type_transition_rates}) then 
the function $\gamma(x)=1/\cosh(x)$.]  

We are primarily interested in the large-$N$ limit, for which it is useful to define two physical quantities that depend on the magnetisation $m$: the mobility $a$ and the free energy $f_{\rm eq}$ which are
\begin{equation}
  \label{eq:activity_free_energy_eq_mfim}
  \fl\qquad \eqalign{
    a(m) &= \frac{1}{2} \gamma(2Jm+h) \sqrt{1-m^{2}}  \\
    f_{\mathrm{eq}}(m) & = - (Jm^{2} + hm) + \frac{1+m}{2}\log \frac{1+m}{2} + \frac{1-m}{2}\log\frac{1-m}{2} \, .
    }
\end{equation}
We identify $a(m)=\lim_{N\to\infty} [N^{-1} w_N(m+2|m) w_N(m|m+2)]^{1/2}$ consistent with the interpretation as a mobility, and the free energy $f_{\rm eq}$ is such that the equilibrium state has a limiting magnetisation distribution $P_{\rm eq}(m)\propto {\rm e}^{-Nf_{\rm eq}(m)}$.  

\subsection{Dynamical free energy}
\label{sec:mfim-Psi}

From \eref{eq:transition_rates_mfim}, one derives the analogue of ${\cal W}_s$, which is a matrix of size $(N+1)\times(N+1)$.  The largest eigenvalue of this matrix is $\Psi_N(s)$.  
Moreover, the function $\psi(s)$ that characterises the large-$N$ limit can be obtained by a simple variational approach.  We take a controlled process analogous to (\ref{eq:transition_rates_opt_process}), in which $\mu$ depends only on $m$.  Due to the mean-field structure of the model, fluctuations are very small and it is sufficient [for the determination of $\psi(s)$] to restrict to $\mu(m) \sim {\rm e}^{-N \Omega(m)}$. Using this ansatz in (\ref{eq:psi-var}) and taking the large-$N$ limit as in (\ref{eq:small-psi}) yields
\begin{equation}\label{eq:free_energy_landau_mfim}
  - \psi(s) = 
   \underset{m \in [-1,1]}{\min} \phi(m,s) \; . 
\end{equation}
with
\begin{equation}
  \label{eq:phi}
    \phi(m,s) = 2 a(m) \left[ \cosh\left( f'_{\rm eq}(m) \right) - e^{-s} \right]
\end{equation}
Here, the prime on $f_{\rm eq}$ indicates a derivative.
Recalling that $-\psi$ is a dynamical free energy, we identify $\phi$ as a Landau-like free-energy density whose minimum gives the true free energy~\cite{garrahan2009first,baek2017dynamical}.  Note however that $\phi$ does not determine the probability to find a configuration with magnetisation $m$ within the biased ensemble.

The next step is to minimise $\phi(m,s)$ over $m$.  %
{Before analysing the complete behavior of $\phi(m,s)$ with respect to the parameters $(s,J,h)$, we shall discuss the equilibrium situation ($s=0$) and the asymptotic regimes ($s\ll -1$, $s\gg 1$). For $s=0$, $\phi(m,s)\geqslant 0$ and the latter is minimized when the thermodynamic force $f_{\rm eq}'(m)$ vanishes, as expected. For $s\gg 1$, $\phi(m,s) \sim 2a(m)\cosh(f_{\rm eq}'(m)) = r(m)$ which is the escape rate \eref{eq:esc-rate} associated with the transition rates \eref{eq:transition_rates_mfim}. Minimising $\phi$ thus reverts to minimizing the escape rate which indeed leads to states with low activity. One should note as well that this low activity state results from a balance between the mobility $a(m)$ and the force $f_{\rm eq}'(m)$ which should both be small.
On the other hand, for $s\ll -1$, $\phi(m,s) \sim -2a(m)e^{-s}$. Minimising $\phi$ thus reverts to maximising the mobility $a(m)$ (hence high activity state). These states do not \emph{a priori} display low forces.
}

To investigate phase transitions in the model, we observe that 
 the Landau free energy $\phi(m, s)$ may be convex in $m$ (with a single minimum) or it can be non-convex,
 depending on the parameters $(J, h, s)$.
 In particular, if $\phi(m,s)$ displays two (degenerate) global minima for a certain value of $s$, one expects to observe phase coexistence and a first-order phase transition.  We now analyse the behaviour of $\phi$ and $\psi$ in some illustrative cases, for the model with Glauber dynamics.  In \sref{sec:phase-mf}, we summarise this information by constructing phase diagrams.

\subsubsection{No magnetic field, $h=0$}
\label{sec:h-zero}
In the absence of a magnetic field, the symmetry of the system under spin-reversal means that $\phi(m,s)=\phi(-m,s)$.  One also sees from the equilibrium free energy in (\ref{eq:activity_free_energy_eq_mfim}) that the system has a classical (thermodynamic) phase transition at $J=J^\ast=0.5$.   This requires that we separate several sub-cases when considering the behavior of $\phi$.

\begin{figure}
  \centering
  \includegraphics[width=0.75\linewidth]{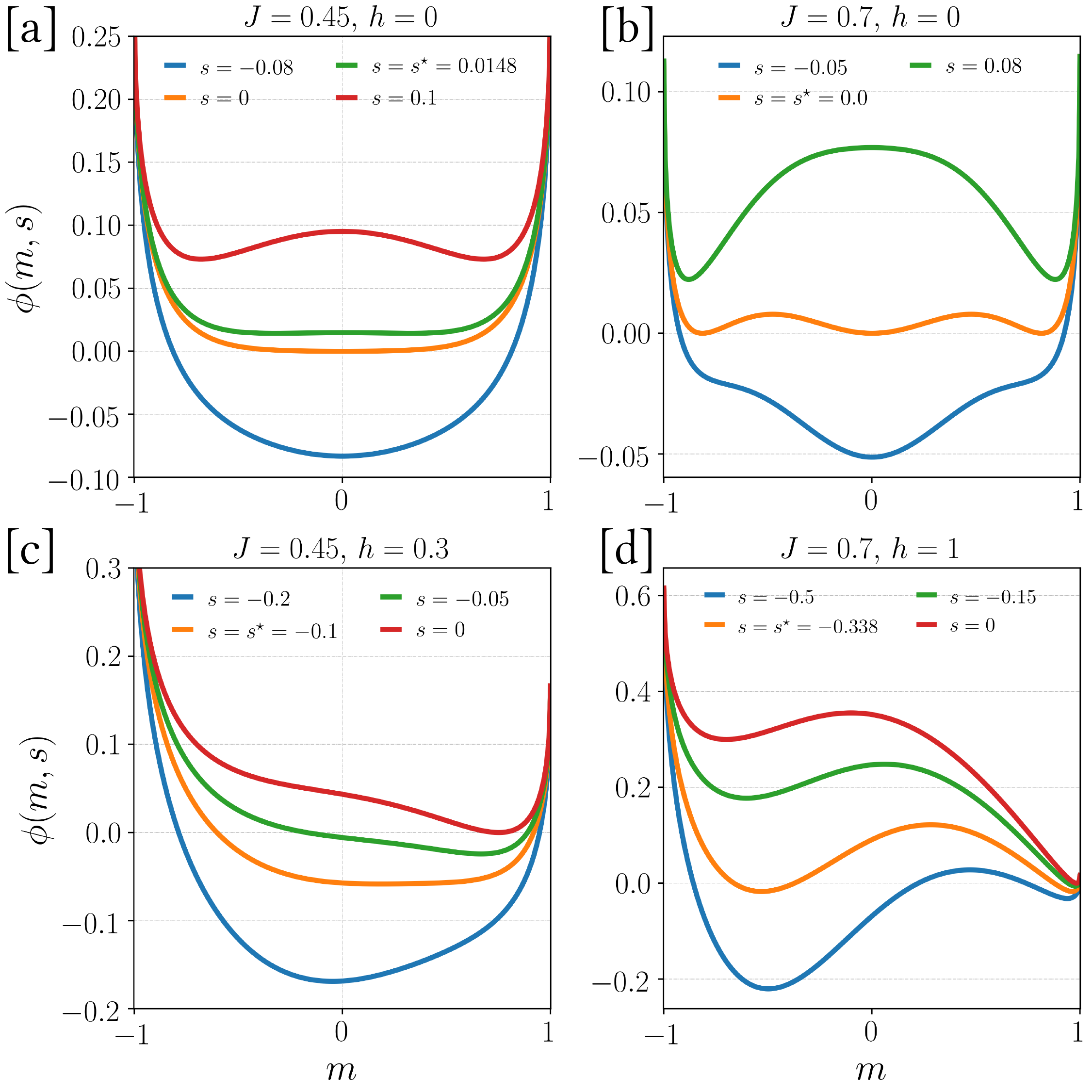}
  \caption{\textbf{Landau free energy $\phi(m,s)$ for different $s$ and $h$}. [\textbf{a}] ($J=0.45 < J^{\ast}$, $h=0$): typical Landau scenario of a second order phase transition at $s=s^{\ast}\simeq 0.0148$. [\textbf{c}] ($J=0.45<J^{\ast}$, $h=0.3$): crossover scenario from $m \lesssim 0$ to $m>0$ at $s\simeq 0.1$. [\textbf{b}] ($J=0.7>J^{\ast}$, $h=0$): first order transition at $s=0$ through a triple point. [\textbf{d}] ($J=0.7>J^{\ast}$, $h=1$): typical Landau scenario of a first order phase transition at $s= s^{\ast} \simeq -0.338$.}
\label{fig:free_energies}
\end{figure}

For $J<J^\ast$, the equilibrium behaviour of the model is paramagnetic, so $\phi(m,0)$ is convex with a minimum at $s=0$.  The behaviour on increasing $s$ is shown in \fref{fig:free_energies}[a].  There is a critical point at some $s=s^\ast$ where $\partial^2 \phi/\partial m^2=0$.  For $s>s^\ast$ then $\phi(m,s)$ has two degenerate minima  (as a function of $m$) corresponding to coexisting states with positive and negative magnetisation.  This is exactly the second-order phase transition 
scenario of Landau \cite{goldenfeld1992lectures} for a critical point at $s=s^\ast>0$ (dependent on $J$).  The physical interpretation is that biasing the system to low activity stabilises the ordered (ferromagnetic) state where the activity is lower.  See also~\cite{lecomte2007thermodynamic} for a similar scenario in mean-field, and~\cite{jack2010large} for the corresponding situation in the one-dimensional Ising model.

For $J>J^\ast$, the situation is more complex, see~\fref{fig:free_energies}[b].  The equilibrium state has two coexisting phases, but $\phi(m,0)$ has \emph{three} minima at $m=0,\pm m^*$, which all have $\phi(m,0)=0$.  For $s>0$, the ferromagnetic states minimise $\phi$ so the behaviour is qualitatively the same as for equilibrium.  However, for $s<0$ the global minimum of $\phi$ is the paramagnetic state $m=0$.  The physical interpretation of this fact is that while $m=0$ corresponds to a local \emph{maximum} of the free energy $f_{\rm eq}$ (and is therefore unlikely at equilibrium), the thermodynamic force $f'_{\rm eq}$ vanishes there.  For small negative $s$, one can minimise $\phi$ in (\ref{eq:free_energy_landau_mfim}) by taking the zero of $f'_{\rm eq}$ with largest mobility $a$.  This corresponds to $m=0$ (because the mobility is maximal there).  In other words, $m=0$ is an unstable fixed point of the (deterministic) mean-field dynamics, and trajectories localised near unstable fixed points can occur with relatively high probability because there are no forces pushing the system away from the fixed point.  See also~\cite{Jack2019-growth}.  

For $J=J^*$ we identify a tricritical point at $(s,h)=(0,0)$
which means in this case that the coefficients of $m^2$ and $m^4$ both vanish in the Taylor expansion of $\phi$, that is $\phi(m,0)\propto m^6 + O(m^8)$.  In this case, small changes in either $J$ or $s$ can lead to large (singular) changes in the energy and/or activity.  We return to this case below.

\subsubsection{Non-zero magnetic field,  $h \neq 0$}
\label{sec:h-nonzero}

In the presence of a magnetic field $h$, the spin-reversal symmetry is broken so $\phi$ is no longer an even function of $m$.  One also finds that $f_{\rm eq}$ has a unique zero in all cases (there is no equilibrium phase coexistence).
The behaviour of the Landau free energy is shown in \fref{fig:free_energies}[c,d] for two representative cases.

For $J<J^\ast$, there are several sub-cases, these are discussed in more detail below.  In \fref{fig:free_energies}[c] one observes a case where the minimum of $\phi$ crosses over smoothly from positive $m$ to negative $m$, as $s$ is reduced from zero.  As anticipated above, this leads to states where the sign of $m$ is opposite to that of $h$, this is the \emph{anomalous} regime  (see also~\cite{van2010second}).  Since we consider Glauber dynamics we have from (\ref{eq:activity_free_energy_eq_mfim}) that the mobility is
\begin{equation}
a(m) = \frac{\sqrt{1-m^2}}{2\cosh(2Jm+h)} \, .
\end{equation}
For $h>0$ we observe that the state of maximal mobility has $m<0$.  The reason is that spins tend to flip more often when the magnetisation is opposite to the magnetic field. From (\ref{eq:free_energy_landau_mfim}) one sees that for large negative $s$ then the minimum of $\phi$ is close to the maximum of $a$.  Hence, the physical origin of the anomalous phase is the fact that $a$ is maximal for some $m$ that is anti-parallel to $h$.  Note that if we had taken dynamics with an exponential rule instead of Glauber rates (leading to $\gamma=1$ in (\ref{eq:transition_rates_mfim}), see \cite{lecomte2007thermodynamic}), then this effect would be absent and the behaviour would be qualitatively different.  In this sense, the phase diagram can depend on details of the model dynamics. 

Figure \ref{fig:free_energies}[d] shows a case with $J>J^{\ast}$, which illustrates a classical first-order phase transition scenario.  The field is positive ($h>0$) so the equilibrium state ($s=0$) corresponds to a global minimum of $\phi(m,0)$ with $m>0$.  There is a secondary (local) minimum at $m<0$.  For positive values of $s$, the large-$m$ state is maintained as the global minimum of $\phi$.  However, on reducing $s$ from zero, the height of the secondary minimum in $\phi$ is reduced.  Eventually a first-order phase transition is reached for some $s=s^*<0$, and the the state with $m<0$ becomes the global minimum.   

\rlj{From (\ref{eq:phi}) \revjg{[with $ J>J^{\ast}, h\neq 0$, see Fig. \ref{fig:free_energies}[d]]}, the secondary minimum in $\phi(m,0)$ is associated with a minimum in the thermodynamic force, due to the non-convex free energy $f_{\rm eq}$. 
The secondary minimum is also associated with a large value of the mobility $a(m)$, compared to the highly-magnetised equilibrium state.  As anticipated in the introduction, these two physical characteristics are expected to be associated with large deviations of high activity, in particular the small thermodynamic force leads to slow relaxation away from this state, which enhances the probability that large-deviation trajectories will be localised there.  For small $h$, the secondary minimum of $\phi$ is located close to the \emph{maximum} of the thermodynamic free-energy $f_{\rm eq}$. This exemplifies a generic mechanism by which large deviations with high activity can be localised near maxima or saddles of the free-energy.  (Note this reasoning is based on $\phi(m,0)$ so it applies only when $|s|$ is not too large, so that the (global) minimum of $\phi(m,s)$ is still close to the local minimum of $\phi(m,0)$.  As $|s|$ gets larger, reasoning based on the natural (unbiased) dynamics of the model becomes less applicable.)
}%

\rlj{To rationalise the first-order transition in this case, note that states between the minimum and the maximum of $f_{\rm eq}$ have large values of $|f_{\rm eq}'|$ and hence large $\phi$.  This suppresses the probability that large-deviation trajectories will visit these states.}   
As a result, the value of $m$ that minimises $\phi$ changes discontinuously as a function of $s$.

\begin{figure}
  \centering
  \includegraphics[width=0.65\linewidth]{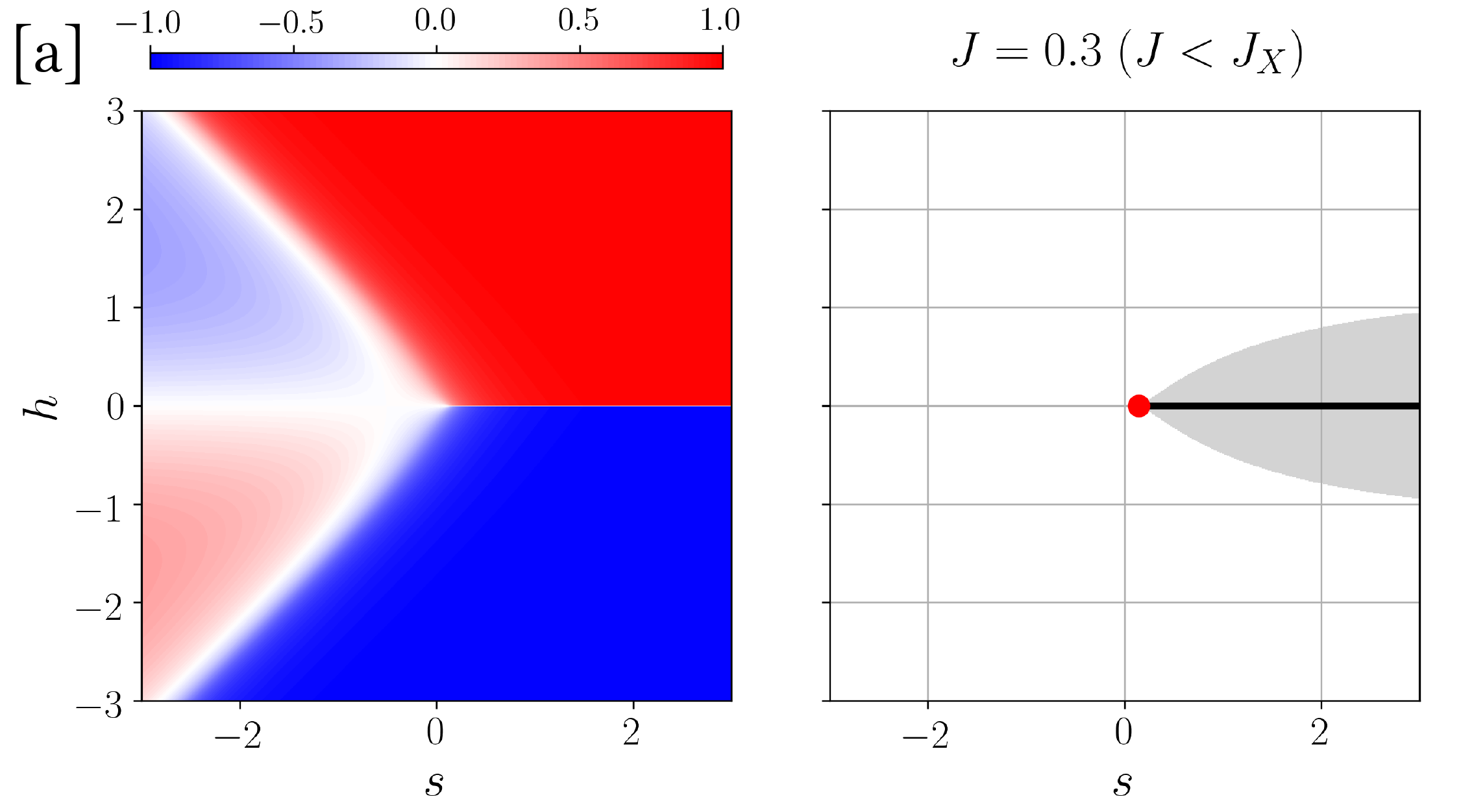}
  \includegraphics[width=0.65\linewidth]{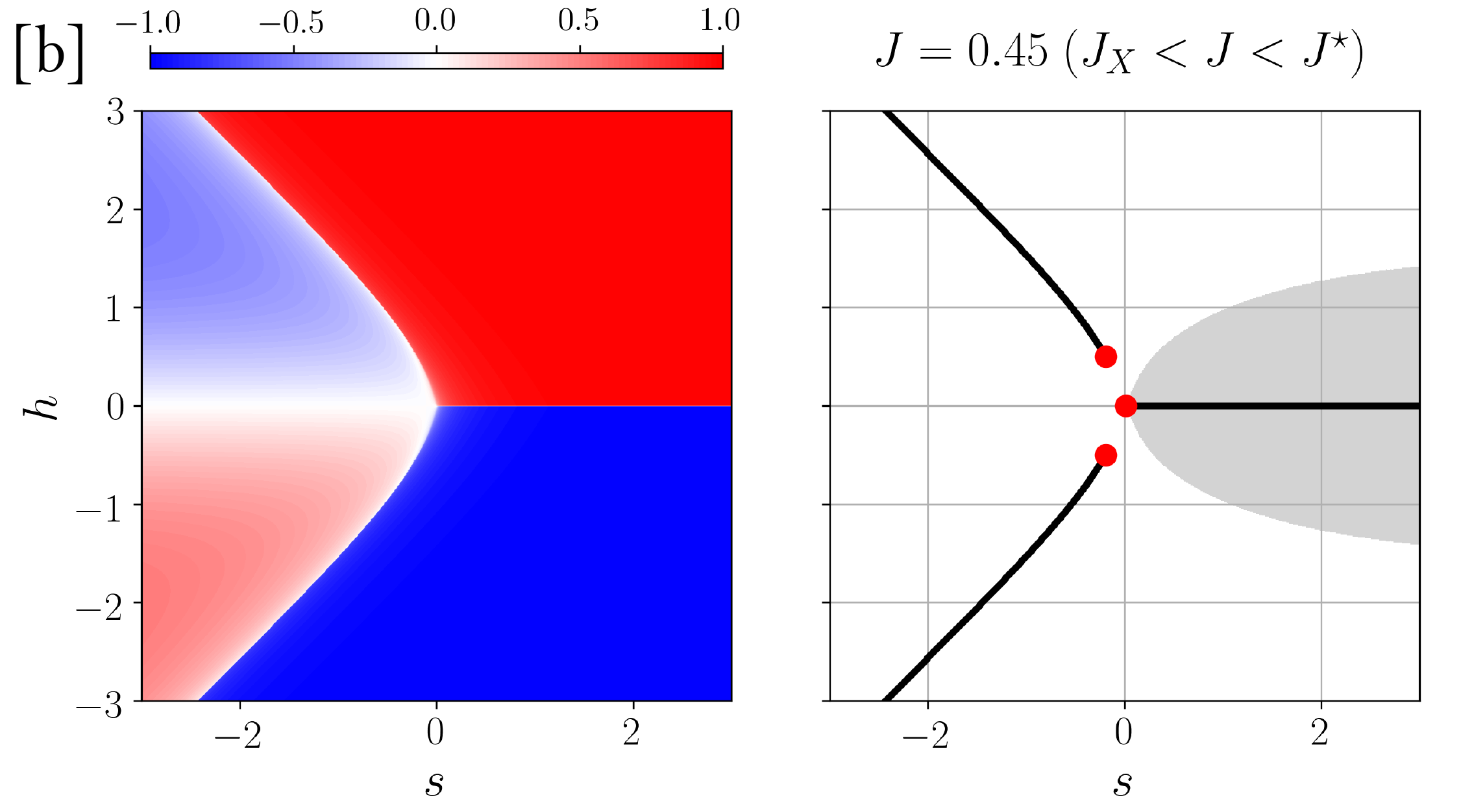}
  \includegraphics[width=0.65\linewidth]{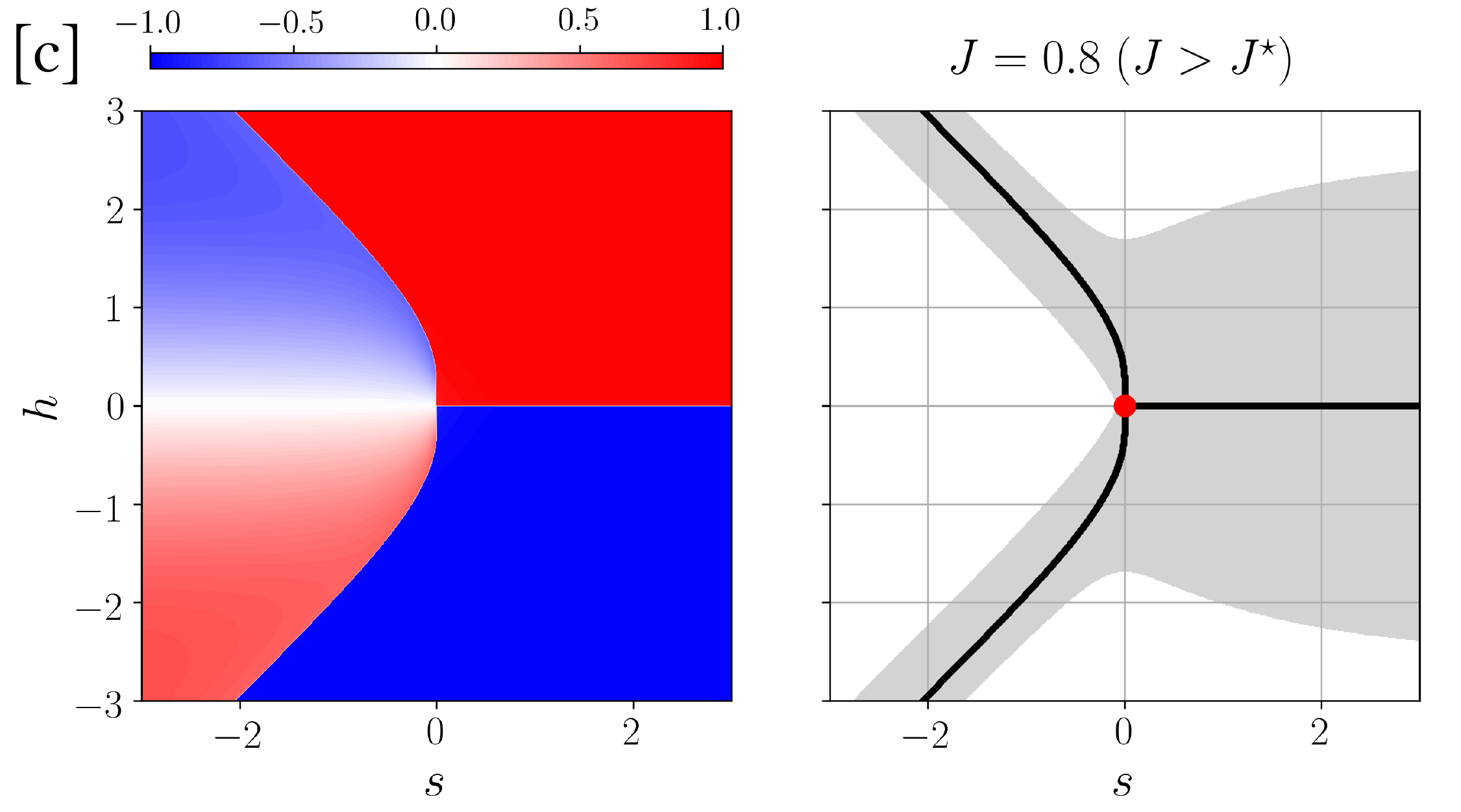}
  \caption{\textbf{Phase diagrams in the $(s, \, h)$ plane of the Mean-Field Ising model}. \emph{Left}: average magnetisation $\avg{m}{s}$ (color) with respect to $(s,h)$. \emph{Right}: non-convex Landau-like free energy regions (grey regions), first order coexistence lines (black lines) and critical points (red dots). [\textbf{a}]: for $J<J_{X}$, the only critical point occurs for $s>0$ (red dot), the behaviour for $s<0$ shows a crossover; [\textbf{b}]: intermediate regime $J_{X}<J<J^{\ast}$ for which each coexistence line ends at a second order critical point (red dots). Non-convex regions for $s<0$ do exist but are very thin (presently hidden by the coexistence black lines). The critical point at $h=0$ is located at $s^{\ast} \simeq 0.0148$ ($J=0.45$). [\textbf{c}]: for $J>J^{\ast}$, all the coexistence lines meet at $(s,h)=(0,0)$ (red dot) which forms a triple point.}
  \label{fig:phase_diag_mfim}
\end{figure}

\subsection{Phase diagrams}
\label{sec:phase-mf}

We now use the form of $\phi(m,s)$ to analyse the phase diagram as a function of $(J,h,s)$.  Results are shown in~\fref{fig:phase_diag_mfim}.

We consider the behaviour of $\phi$ as a function of $m$ (at constant $s$).  If $\phi(m,s)$ has a unique global minimum with $\partial^2\phi/\partial m^2>0$ then the system has a single phase.  If there is a unique global minimum with $\partial^2\phi/\partial m^2=0$ then the system is at a critical point.  (There may also be tricritical points where higher derivatives of $\phi$ also vanish.)  If the global minimum is not unique then the system lies on a first-order transition line.    On varying $(h,s)$, we find three kinds of behaviour, depending on the value of $J$.  We already identified $J^\ast=0.5$ as the ferromagnetic critical coupling for the equilibrium model. We also identify a crossover at {$J=J_X\approx0.402964$ [see \ref{sec:app:computation_J_cross}} for its derivation], whose meaning is discussed below.  Regions of the phase diagram where $\phi$ has multiple local minima are shaded in~\fref{fig:phase_diag_mfim}.  To the extent that $\phi$ is a Landau free energy, local minima can be interpreted as metastable phases.  However, we will see in \sref{sec:path-int} that this interpretation requires some care.

For $J<J_X$ and varying $(h,s)$, there is a single critical point at $(0,s^*)$ with $s^*>0$, see \fref{fig:phase_diag_mfim}a.  This critical point is the one identified in Sec.~\ref{sec:h-zero}, where positive $s$ (low activity) acts to promote ferromagnetic order, as in~\cite{lecomte2007thermodynamic}.

For $J>J^*$, the equilibrium behaviour is ferromagnetic and $(s,h)=(0,0)$ is a triple point where $\phi$ has three degenerate minima.  The $(s,h)$ plane contains three first-order transition lines, which all meet at the triple point: see \fref{fig:phase_diag_mfim}b.   For $s>0$ there is a first-order transition line at $h=0$ and the magnetisation of the system is discontinuous across this line, with $m$ having the same sign as $h$.  In this sense the behaviour for $s>0$ is the same as that for $s=0$.  For $s<0$ one observes the first-order phase transition discussed in \sref{sec:h-nonzero} which separates the equilibrium ferromagnetic phase from an anomalous phase where the magnetisation has the opposite sign to $h$.

For the intermediate case $J_X<J<J^*$ the equilibrium behaviour is paramagnetic but the system has three critical points and three first-order lines where phase coexistence takes place.  For $s>0$ (low activity), the behaviour is similar to $J<J_X$ with a single critical point at $s_c>0$.  However, for $s<0$ (high activity) one again observes of anomalous phases which may coexist with regular (paramagnetic) phases.  
The coupling $J_X$ is the value that separates whether the system has three critical points as in \fref{fig:phase_diag_mfim}[b] or only one as in \fref{fig:phase_diag_mfim}[a]. {More precisely, $J_{X}$ is the coupling at which the critical point is sent to $(s,h)\to(-\infty,\infty)$ such that $\Delta = s+h$ stays finite (see \ref{sec:app:computation_J_cross} for more details)}. 
On the other hand, as $J\to J^\ast$ (from below), the three critical points all approach the point $(s,h)=(0,0)$, this becomes a tricritical point for $J=J^\ast$.  
This concludes our analysis of the dynamical phase behaviour of the mean-field Ising model.

\subsection{Effective dynamics in the large system size: effective force and quasi-potential}\label{sec:effective_dyn_mfim}
\label{sec:path-int}

We have computed the dynamical phase diagram of the mean-field Ising model using a variational characterisation of the largest eigenvalue of the operator ${\cal W}_s$.  This amounts to minimising the function $\phi$, which determines the dominant value of the magnetisation $m$, within the biased ensemble.  However, the mean-field aspect of the model allows a more detailed characterisation of trajectories within the biased ensemble.  In particular, it is possible to compute the distribution of $m$, and the dominant paths by which rare values are visited.  These considerations are particularly relevant at points of dynamical phase coexistence, as we now discuss (see also~\cite{nemoto2014finite,nemoto2017finite,JackNemoto2019}).

\subsubsection{Path integral formulation}

It is useful to consider a path-integral formulation of the dynamics, following Martin-Siggia-Rose-De Dominicis-Jensen (MSRJD) \cite{de1978field,PhysRevA.8.423,Janssen1976,PhysRevB.18.4913}.
For large $N$, this amounts to writing the path probability density $P_s(\Theta_{\tobs})$ of the biased ensemble as \cite{lefevre2007dynamics,andreanov2006field,thompson2011lattice,Ge2017,Bouchet2016} 
\begin{equation}\label{eq:msrjd_mfim}
  P_s(\Theta_{\tobs}) \underset{N\to\infty}{\sim} \frac{p_{0}(m_0)}{Z(s,T)}  \int \, \mathcal{D} \hat{m} \,  {\rm e}^{-N \mathcal{S}_{s}[m,\hat{m}] }
  \, ,
\end{equation}
where 
\begin{equation}
\mathcal{S}_{s}[m,\hat{m}]  = \int_{0}^{\tobs} \! \mathrm{d}t  \, \left\{ \hat{m}(t) \dot{m}(t) - H_s\left(m(t), \hat{m}(t)\right)   \right\}
\end{equation}
and $p_{0}(m_0)$ is the probability of the initial condition.  
The Hamiltonian $H_s$ may be derived as
\begin{equation}
\fl \qquad
\label{eq:hamiltonian_mfim}
    H_s(m, \hat{m}) 
    \equiv \lim_{N\to\infty} \frac{1}{N} \lim_{\mathrm{d}t \to 0} \frac{ \avg{ {\rm e}^{\hat{m} N\left[ m(t+\mathrm{d}t) - m(t) \right] - s{\cal A}(t,t+\mathrm{d}t)} \middle | m(t) = m}{} - 1}{\mathrm{d}t} 
\end{equation}
where the notation $\avg{ \cdot{} \middle| \cdot{} }{}$ indicates a conditional average and ${\cal A}(t,t+{\rm d}t)$ is the contribution to the dynamical activity for the time interval $[t,t+{\rm d}t]$ (for example, the number of spin flips in this interval).  
Note that (\ref{eq:msrjd_mfim}) involves a limit of large-$N$ at fixed $T$.  
Also
\begin{equation}
  \label{eq:partition_function_mfim}
  Z(s,\tobs) \underset{ N \to \infty }{ \sim } \int \mathcal{D}m \, \mathcal{D}\hat{m} \,  p_{0}(m_0) e^{- N \mathcal{S}_{s}[m,\hat{m}] }  \, .
\end{equation}
From the dynamical rules of the model and taking ${\cal A}={\cal N}$, one finds
\begin{equation}
H_s = 2a(m) \left\{ {\rm e}^{-s} \cosh\left[ 2\hat{m} - f'_{\rm eq}(m) \right] - \cosh f'_{\rm eq}(m) \right\} \; .
\label{eq:biased_hamiltonian_mfim}
\end{equation}

This path-integral formalism is convenient because integrals such as (\ref{eq:partition_function_mfim}) can be computed by saddle-point methods, thanks to the large parameter $N$ appearing in the exponent.  Moreover, the action has an Hamiltonian structure which means that given two times $t_0,t_1$ and two points $(m_0,m_1)$, the most likely path (instanton) connecting these points has a constant value of the Hamiltonian $H_s$.  This is easily verified via the Euler-Lagrange equations for the action $\mathcal{S}_{s}$ which are
\begin{equation}
  \label{eq:hamilton_eq_mfim}
  \eqalign{
    \dot{m}(t)  & = \pdev{H_{s}}{\hat{m}}{}(m(t), \hat{m}(t)) \\
    \dot{\hat{m}}(t) & = - \pdev{H_{s}}{m}{}(m(t), \hat{m}(t)) \; ,
    } 
\end{equation}
  It follows that the most likely path with $m(t_0)=m_0$ and $m(t_1)=m_1$ has $H_{s}(m(t), \hat{m}(t))$ independent of $t$ (for $t_0\leq t\leq t_1$).  

As a first consequence of this observation, we describe an alternative derivation of (\ref{eq:free_energy_landau_mfim},\ref{eq:phi}).  
Consider stationary trajectories where both $m$ and $\hat{m}$ are independent of $t$.  The action depends only on $ H_s(m,\hat{m})$ and the stationary trajectory with maximal probability is obtained by maximising this quantity over $\hat{m}$.  The maximum occurs at
\begin{equation}
  \label{eq:mhat_stationary}
    \hat{m}^{\ast}(m)  = \frac{1}{2}f_{\mathrm{eq}}^{\prime}(m) 
\end{equation}
Hence, comparing \eref{eq:biased_hamiltonian_mfim} with \eref{eq:phi} one sees that
\begin{equation}
H_s(m,\hat{m}^{\ast}(m)) = - \phi(m,s) 
\end{equation}
so that the action for such a path is ${\cal S}_s=-NT\phi(m,s)$.
Hence, assuming that the integral in \eref{eq:partition_function_mfim} is dominated by such trajectories and using (\ref{eq:small-psi}) recovers \eref{eq:free_energy_landau_mfim}.\footnote{%
We have only sketched the relevant argument here.  The assumption that a single trajectory dominates in  \eref{eq:partition_function_mfim} is justified by the large-$N$ limit.  (There are exceptions at points of phase coexistence but $\psi$ is continuous so these isolated points do not pose a problem.)  The assumption that the dominant trajectory is stationary is an approximation.
To evaluate  \eref{eq:partition_function_mfim} one should consider non-stationary trajectories with transient behaviour close to $t=0$ and $t=T$.  However, the large-$T$ limit in (\ref{eq:small-psi}) means that these transient regimes can be neglected for the computation of $\psi$.}

\begin{figure}
\hspace{2cm}\includegraphics[width=5cm]{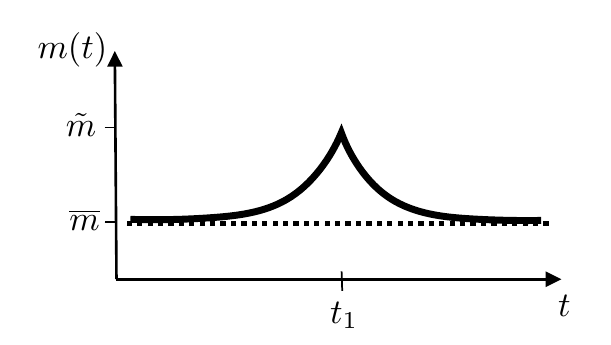}
\caption{Sketches of paths that contribute to the integrals in (\ref{eq:Omega-path}), for systems away from phase coexistence.  The dotted line shows the stationary path that dominates the partition function $Z$.  (Transient regimes near $t=0,T$ are not shown.)  The solid line illustrates the instanton path that dominates the numerator of (\ref{eq:Omega-path}).  It makes an excursion from $\overline{m}$ to $\tilde{m}$ before relaxing back to $\tilde{m}$.  Its derivative is discontinuous at $t_1$ but its Hamiltonian is constant throughout.}
\label{fig:inst}
\end{figure}

\subsubsection{Quasi-potential and instanton paths --- systems away from phase coexistence}
\label{sec:quasipot_instanton}

Recall from \sref{sec:mfim-Psi} that the stationary probability distribution within the biased ensemble is $\mu^*(m) \sim {\rm e}^{-N \Omega_s(m)}$.  The quantity $\Omega_s(m)$ is called the quasi-potential \cite{Bertini2015,freidlin1998random}, by analogy with a Boltzmann distribution based on a potential energy (indeed for $s=0$ then $\Omega_s=f_{\rm eq}- \min f_{\rm eq}$).  

The variational principle (\ref{eq:psi-var}) allows computation of $\mu^*$ and hence of $\Omega_s$.  Here we compute $\Omega_s$ from the path integral, as
\begin{equation}
\fl \qquad
\label{eq:Omega-path}
\Omega_s(\tilde{m}) = \lim_{N\to\infty} \lim_{T\to\infty} \frac{-1}{N} \log \frac{ \int \mathcal{D}m \, \mathcal{D}\hat{m} \,  p_{0}(m_0) \delta(m(t_1)-\tilde{m}) e^{- N \mathcal{S}_{s}[m,\hat{m}] }  \, .
 }{ Z(s,\tobs) }
\end{equation}
where $1\ll t_1\ll T$ (the result is independent of $t_1$, in that regime).  Note that the limit of large-$T$ is taken \emph{before} the limit of large $N$, we return to this point below.

Let us first restrict to situations away from phase coexistence, so $\phi(m,s)$ has a unique (global) minimum at $m=\overline{m}$.  This is the typical magnetisation in the biased ensemble so $\Omega_s(\overline{m})=0$.  
The key insight is that the probability to find magnetisation $\tilde m$ within the biased ensemble is controlled by an instanton that begins at $m=\overline{m}$, makes an excursion  to $\tilde m$, and then relaxes back to $\overline{m}$, see \fref{fig:inst}.  The instanton minimises the action ${\cal S}_s$, subject to this constraint. The integral in the numerator of (\ref{eq:Omega-path}) is dominated by the instanton path and the partition function in the denominator is controlled by the stationary path described above.  Hence $\Omega_s$ depends only on the difference in action between these paths.
To ensure that the action is minimal as $T\to\infty$, the two paths must have the same value of the Hamiltonian which is
\begin{equation}
H_s(m(t),\hat{m}(t)) = -\phi^*
\end{equation}
From (\ref{eq:biased_hamiltonian_mfim}),  the instanton can be characterised by finding $\hat{m}$ parameterically as a function of $m$. One finds that $\hat{m}(t) = \Lambda_s^\pm(m(t))$ with 
\begin{equation}
\label{eq:Lambda}
\Lambda_s^\pm(m) = \frac12 \left[ f'_{\rm eq}(m) \pm \arcosh\left( 1 +  \frac{ {\rm e}^s }{ 2a(m) }   [\phi(m,s)-\phi^* ]\right) \right] \,
\end{equation}
such that $\Lambda_{s}^{\pm}(\overline{m}) = 0$. We have also
\begin{equation}
\label{eq:dot-m}
\dot{m} = 4 a(m) {\rm e}^{-s} \sinh( 2\hat{m} - f'_{\rm eq}(m) ) \; .
\end{equation}
Eqs.~(\ref{eq:Lambda}, \ref{eq:dot-m}) are sufficient to construct the instanton.  From \eref{eq:dot-m}, the sign in $\Lambda_s^\pm$ indicates whether $m$ is increasing or decreasing as a function of time.  The instanton then requires that we combine the two solutions $\Lambda_s^+$ and $\Lambda_s^-$.
For $\tilde{m}>\overline{m}$, one takes 
$\hat{m}(t) = \Lambda_s^+(m(t))$ for $t<t_1$ and $\hat{m}(t) = \Lambda_s^-(m(t))$ for $t>t_1$.  The opposite case holds for $\tilde{m}<\overline{m}$. 

\begin{figure}
  \centering
  \includegraphics[width=1.\linewidth]{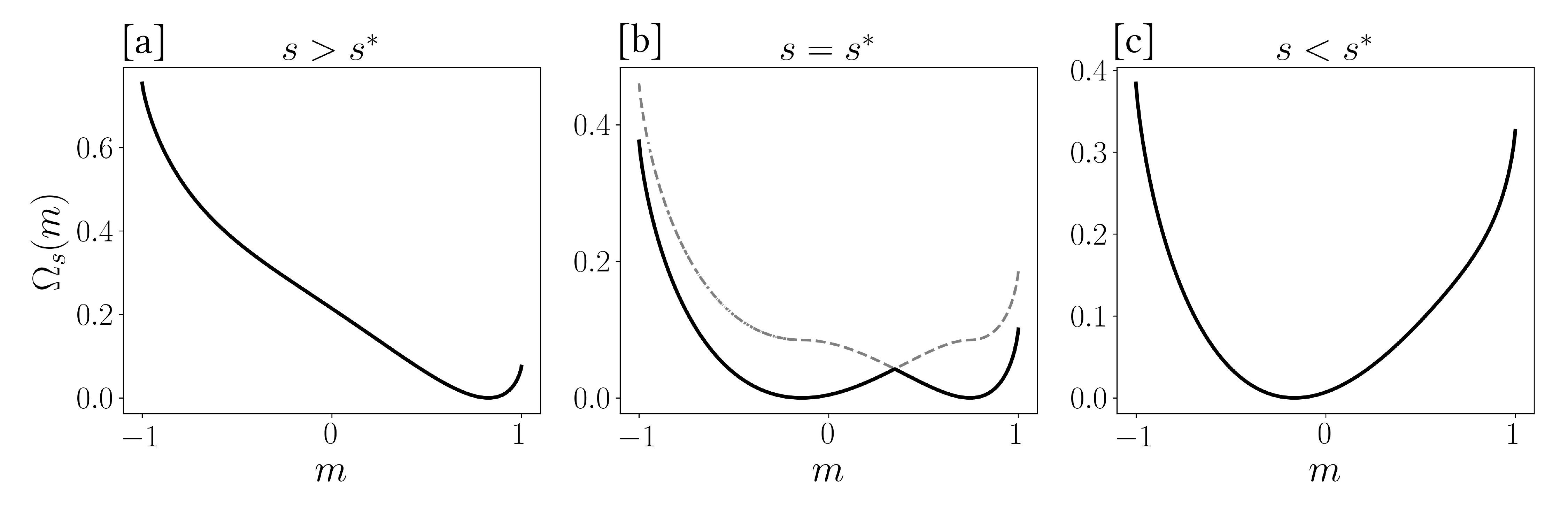}
  \caption{\textbf{Quasi-potential of the Mean-Field Ising Model}. Parameters are: $J=0.5$, $h=0.5$. [\textbf{a}]: $s=-0.1>s^{\ast}$; [\textbf{b}]: $s=s^{\ast} \approx -0.16742 $; [\textbf{c}]: $s=-0.2<s^{\ast}$. One can observe at $s=s^{\ast}$ the two branches $\Omega_{s, m_{\mathrm{i}}^{\ast}}$ and $\Omega_{s, m_{\mathrm{a}}^{\ast}}$ associated with the active and inactive phases (grey dashed lines).}
  \label{fig:quasipotential_mfim}
\end{figure}

Returning to (\ref{eq:Omega-path}), the integral is done by the saddle-point method and one uses that both paths in \fref{fig:inst} have the same value of the Hamiltonian to write
\begin{equation}
\label{equ:Om1}
\Omega_s(\tilde{m})  = \int_0^T \Lambda_s^\pm(m(t)) \dot{m}(t) \mathrm{d}t ,
\end{equation} 
where the integral is evaluated along the instanton.  For $t<t_1$ the relevant path goes monotonically from $\overline{m}$ to $\tilde{m}$, after which it returns monotonically to $\overline{m}$.  Assuming $\tilde{m}>\overline{m}$ and  changing the integration variable in (\ref{equ:Om1}) yields
$ 
\Omega_s(\tilde{m}) = \int_{\overline{m}}^{\tilde{m}} \Lambda_s^+(m) \mathrm{d}m + \int^{\overline{m}}_{\tilde{m}} \Lambda_s^-(m) \mathrm{d}m
$ 
with a similar expression for $\tilde{m}<\overline{m}$.  Finally one obtains an explicit formula for the quasipotential
\begin{equation}
\label{equ:Om-final}
\Omega_s(\tilde{m}) =  \left| \int_{\overline{m}}^{\tilde{m}} \arcosh\left( 1 +  \frac{ {\rm e}^s }{ 2a(m) }   [\phi(m,s)-\phi^* ]\right) \mathrm{d}m  \right|  \; .
\end{equation}
Note, for $s=0$ one has $\phi^*=0$ and one recovers $\Omega_s(m)=f_{\rm eq}(m)-f_{\rm eq}(\overline{m}) $, as it must be.
The behaviour of the quasipotential is shown in Fig.~\ref{fig:quasipotential_mfim}[a,c], for points in the active and inactive phases.  The case of phase coexistence is different and will be discussed in the next section.

We observe that the integrand is non-negative in \eref{equ:Om-final}, which means that the derivative of $\Omega_s$ is non-negative for $\tilde{m}>\overline{m}$, and non-positive for $\tilde{m}<\overline{m}$  (excluding systems at phase coexistence).  In particular, this means that $\Omega_s$ has exactly one minimum (at $\overline{m}$).  This contrasts with the variational free energy $\phi$ which can have local minima.  This has consequences for metastability in biased ensembles, see \sref{sec:qp-discuss}.

\subsubsection{Quasi-potential and instanton paths for systems at phase coexistence}

\begin{figure}
\hspace{2cm}\includegraphics[width=9cm]{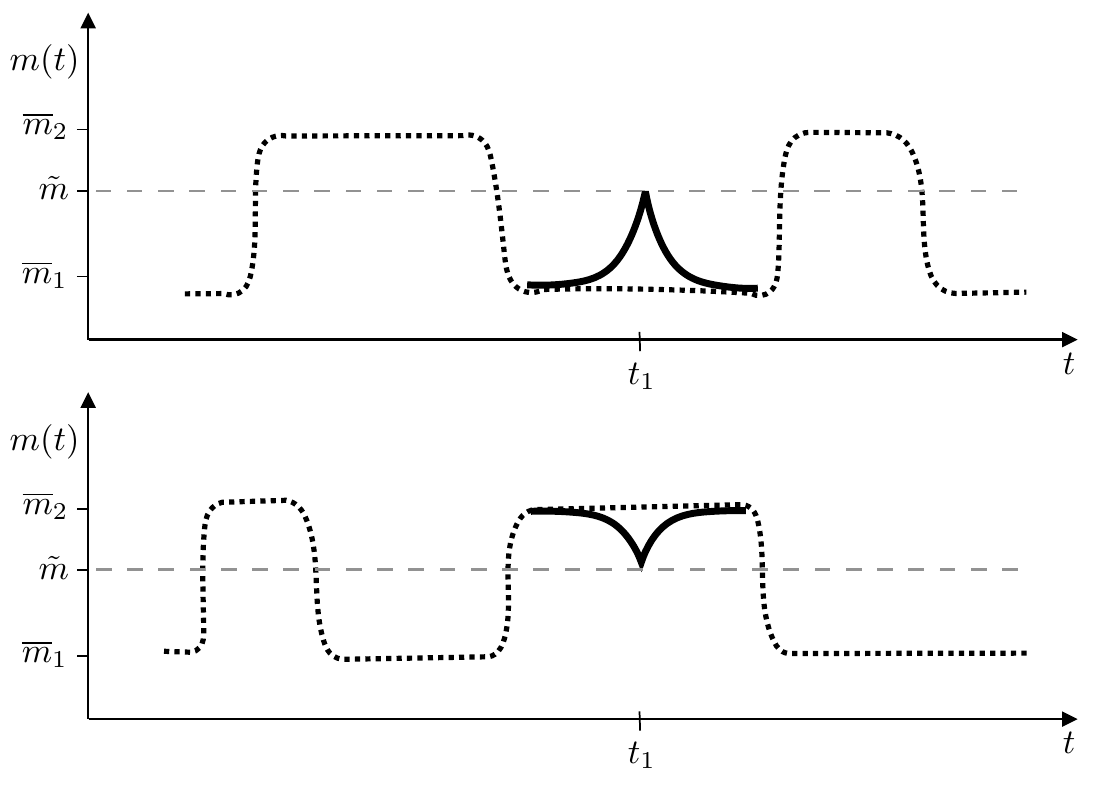}
\caption{Sketches of paths that contribute to the integrals in (\ref{eq:Omega-path}), for a system at phase coexistence, with very large $T$ and finite $N$.  The dotted lines show typical paths within the biased ensemble.   Each path visits both phases.  For every such path in the biased ensemble, there is a corresponding instanton path that makes an excursion to $\tilde{m}$ at time $t_1$.
The excursions are shown with solid lines; the instantons coincide with the dotted lines for other values of $t$.  The two possible instantons that lead to $\tilde{m}$ correspond to the two branches in fig.~\ref{fig:quasipotential_mfim}, the quasipotential is determined by the branch with the smaller action. }
\label{fig:inst2}
\end{figure}

For systems at dynamical phase coexistence, the variational free energy $\phi$ has two (or more) minima, which are the coexisting phases.  This introduces several subtle aspects when evaluating the integral in \eref{eq:Omega-path}.   We restrict to  the case where two phases coexist, with magnetisations $\overline{m}_{1,2}$. (The extension to multiple phases is straightforward.)  In \eref{eq:Omega-path}, the limit of large-$T$ is taken before the limit of large-$N$, which means that there are many paths contributing to the partition function $Z(s,T)$ --- a typical path visits both  phases, making many transitions between them.  The structure of typical paths is shown in \fref{fig:inst2} with dotted lines.  The time spent between transitions is of order ${\rm e}^{N\Omega^*}$, the determination of the barrier height $\Omega^*$ will be discussed below.  The important observation is that we take $T\to\infty$ at finite $N$ so the number of transitions in a typical path is of order $T{\rm e}^{-N\Omega^*}$ which diverges in the limit.  All these paths have the same value for the Hamiltonian, which is $\phi^*$.\footnote{%
Since trajectories visit both phases, we find that properties of the biased ensemble are independent of the initial conditions $p_0$ used in its definition.  This might not be the case if one took the large-$N$ limit \emph{before} the large-$T$ limit in (\ref{eq:Omega-path}), because the probability to reach $\tilde{m}$ by an excursion from $\overline{m}_1$ would depend on the probability that the initial condition comes from that metastable state.
}

To evaluate the quasipotential, observe from \fref{fig:inst2} that the instanton can make its excursion to $\tilde{m}$ from either phase, after which it returns to the same phase.  (Cases where the instanton starts its excursion from one phase and ends in the other will be discussed below.)  One may then repeat the analysis leading to (\ref{equ:Om-final}), noting that while there are many paths contributing to both the numerator and denominator of \eref{eq:Omega-path}, these paths are in one-to-one correspondence. For each corresponding pair, the difference in action is given by an integral similar to
the right hand side of (\ref{equ:Om1}).  Since paths of minimal action dominate the integrals in \eref{eq:Omega-path} one then finds
\begin{equation}
\label{equ:Om-coex}
\Omega_s(\tilde{m}) = \min_{\alpha=1,2} \left| \int_{\overline{m}_\alpha}^{\tilde{m}} \arcosh\left( 1 +  \frac{ {\rm e}^s }{ 2a(m) }   [\phi(m,s)-\phi^* ]\right) \mathrm{d}m  \right|  \; .
\end{equation}
That is, the quasipotential is obtained by minimising the action over instantons that may start in either phase.  Fig.~\ref{fig:quasipotential_mfim}[b] shows an example of a quasipotential that is obtained in this way.  
This construction leads naturally to a quasipotential whose derivative is discontinuous at some $m=m^\ddag$ between $\overline{m}_1$ and $\overline{m}_2$.  This $m^\ddag$ corresponds to a transition state and we identify the barrier height (defined above) as $\Omega^*=\Omega_s(m^\ddag)$.  
For $s=0$, $\phi^{\ast}=0$ and $\phi(m,0)=2a(m)\left(\cosh(f_{\rm eq}'(m)) - 1 \right)$ [see \eref{eq:phi}]. Hence the quasi-potential \eref{equ:Om-coex} becomes $\Omega_{0}(\tilde{m}) = f_{\rm eq}(\tilde{m}) - \min f_{\rm eq}(\overline{m}_{\alpha})$ as expected.

Having characterised the barrier, it is natural to consider instantons which start at $\overline{m}_1$ and end at $\overline{m}_2$.  Assuming that these pass through $m^\ddag$ at time $t_1$, the relevant paths can be obtained by combining the instanton from $\overline{m}_1$ to $m^\ddag$ (taking the part with $t<t_1$) and the instanton from $\overline{m}_2$ (taking the part with $t>t_1$).  These two instantons have the same value of the Hamiltonian.  The physical consequence of the discontinuity in $\Omega_s'$ at $m^\ddag$ is that the top of the barrier does not correspond to a fixed point of Hamilton's equations, which means that the instanton passes through the barrier with finite velocity $\dot{m}$.  This is distinct from the equilibrium (zero-bias, $s=0$) case where $\Omega_{0}'(m^{\ddag}) = 0$.

We make one further comment about dynamical phase coexistence.  We have emphasised that the rate for transitions between the coexisting phases in the biased ensemble scales as $\omega_{0} = {\rm e}^{-N\Omega_{s^{\ast}}(m^{\ddag})}$.  As in~\cite{JackNemoto2019}, the trajectories dominating the path integral can then be described by a Poisson process where the system hops between the phases with this rate.  Since these two phases have different values of the dynamical activity (recalling that the activity is proportional to $N$, we denote these by $Na_1,Na_2$), the dominant contribution to fluctuations of the time-averaged activity can be captured by this Poisson process.   In particular (see \ref{app:poisson}), this simple model leads to a crossover function for the dynamical free energy (valid for $s$ very close to $s^*$) is
\begin{equation}
\fl\qquad
\Psi_{N}(s) \approx -(s-s^{\ast})  \frac{N(a_1 + a_2)}{2} - \omega_0 + \sqrt{\frac{N^2}{4}(s-s^{\ast})^{2}(a_1 - a_2)^{2} +\omega_0^2}
\label{eq:finite-size-psi}
\end{equation}
and hence
\begin{equation}
  \label{eq:finite_size_scaling_ddpsi_sstar_mfim}
  \Psi_{N}''(s^{\ast}) \sim N\frac{{(a_1 - a_2)}^{2}}{4}e^{N\Omega_{s^{\ast}}(m^{\ddag})} \, .
\end{equation}
That is, the curvature of the free energy (and hence the derivative of the order parameter) diverges exponentially with system size.
In fact, this mapping of systems at phase coexistence to a Poisson process is very general \cite{nemoto2017finite,JackNemoto2019}, we will come back to these results when considering the one-dimensional Ising model in \sref{sec:Ising1d}.

\subsubsection{Discussion --- quasi-potential and variational free energy}
\label{sec:qp-discuss}

We have computed two functions $\phi$ and $\Omega_{s}$ which quantify probabilities in the biased ensemble.  That is, $\phi(m,s)$ corresponds to the log-probability of a long trajectory where the magnetisation is $m$ for (almost) all times $t$.  On the other hand, $\Omega_s(m)$ corresponds to the log-probability that $m(t)=m$ at a single time $t$ (far from initial and final times).

\revjg{Note also that the optimally-controlled dynamics that reproduces the trajectories of the biased ensemble can be obtained (for large $N$) by adding a control potential $ U^{\rm con}(m)=[\Omega_s(m)-f_{\rm eq}(m)]$, as in Eq. \eref{eq:optimal_control_forces_def}.  The control force is the gradient of the potential, $[f_{\rm eq}'(m)-\Omega_s'(m)]$.
In the stationary state $\bar{m}$ of the controlled dynamics, one has $\Omega_{s}'(\bar{m})=0$ so the control force is simply $f_{\rm eq}'(\bar{m})$.}
\rlj{For high-activity phases when $|s|$ is not too large, we have explained that the optimal-control forces act to localise the system near a maximum of the free energy.  The thermodynamic force $f_{\rm eq}'$ is small there, so relatively weak control forces are enough to accomplish this.  (It is a general result that large-deviation mechanisms tend to have weak control forces~\cite{chetrite2015variational,Jack2019}.)}

The functions $\phi$ and $\Omega_s$ are both minimal at $m=\overline{m}$ and both have features that resembles a Landau free-energy in equilibrium phase transitions.  However, we emphasise that the relevant probabilities are qualitatively different, and the functions have different forms.  
For example, in systems away from phase coexistence then $\phi$ may have local minima, but $\Omega_{s}$ has a single minimum.
At phase coexistence, both $\phi$ and $\Omega_{s}$ have two minima, but $\phi$ is a smooth function while $\Omega_{s}$ has a discontinuity in its derivative at $m=m^\ddag$.  

Physically, the important point is that local minima of $\Omega_s$ would correspond to metastable states of the optimally-controlled system (these would be states for which equilibrating the optimally-controlled system would require a time that diverges exponentially with $N$). However, such states do not appear in our analysis: Phase coexistence may occur at some $s=s^*$ but there are no metastable states that survive on perturbing $s$ away from $s^*$.  On the other hand, local minima of $\phi$ do survive for $s\neq s^*$, but these have a different physical interpretation --- they correspond to stationary trajectories at magnetisation $m$ for which the probability decreases on perturbing $m$ away from $\tilde{m}$.  This can be interpreted as a kind of metastability \emph{in trajectory space}, in that homogeneous perturbations to the trajectory act to increase the dynamical free energy \cite{garrahan2009first}.  This behaviour is quite different from classical (thermodynamic) metastability which occurs in configuration space, and describes the local stability of configurations (or thermodynamic states) to homogeneous perturbation.  Of course, thermodynamic metastability has dynamical implications; the point here is that metastability in trajectory space is distinct from thermodynamic metastability, and has a different set of implications for dynamical behaviour.  The biased ensemble of trajectories may exhibit metastability in trajectory space but there is no thermodynamic metastability in the optimally-controlled system (because $\Omega_s$ does not have local minima).

\section{1D Ising model in a magnetic field}\label{sec:Ising1d}

We now consider the Ising model in $d=1$.  
This model does not have any equilibrium phase transitions, but dynamical phase transitions are still present~\cite{jack2010large,loscar2011thermodynamics,vasiloiu2019trajectory}.  We will find that the behaviour of the 1D model (for positive $J$) resembles that of the mean-field model for $J<0.5$. 

For $h=0$, exact results are available, 
based on a mapping of a quantum-Ising chain~\cite{garrahan2009first,jack2010large}.  Details are given in \ref{app:exact_ising1D} which also corrects two small errors in \cite{jack2010large}.  There is a critical point at $(h,s)=(0,s_c)$ with $s_{c}=-\ln \tanh(2\beta J)$, see \eref{eq:app:s_star_exact_h0} from \ref{app:exact_ising1D}.  For $h\neq0$ we are not aware of any exact solution so we use instead numerical methods based on exact diagonalisation (for small systems) and the cloning algorithm~\cite{giardina2006direct, lecomte2007numerical}.
At this point we also recall that~\cite{loscar2011thermodynamics} considered similar large deviations to those discussed here, including the case of $h\neq0$.  However, their analysis was restricted to $s>0$ so they did not consider the anomalous regime where $m$ is antiparallel to $h$.  

\begin{figure}
  \centering
  \includegraphics[width=0.7\linewidth]{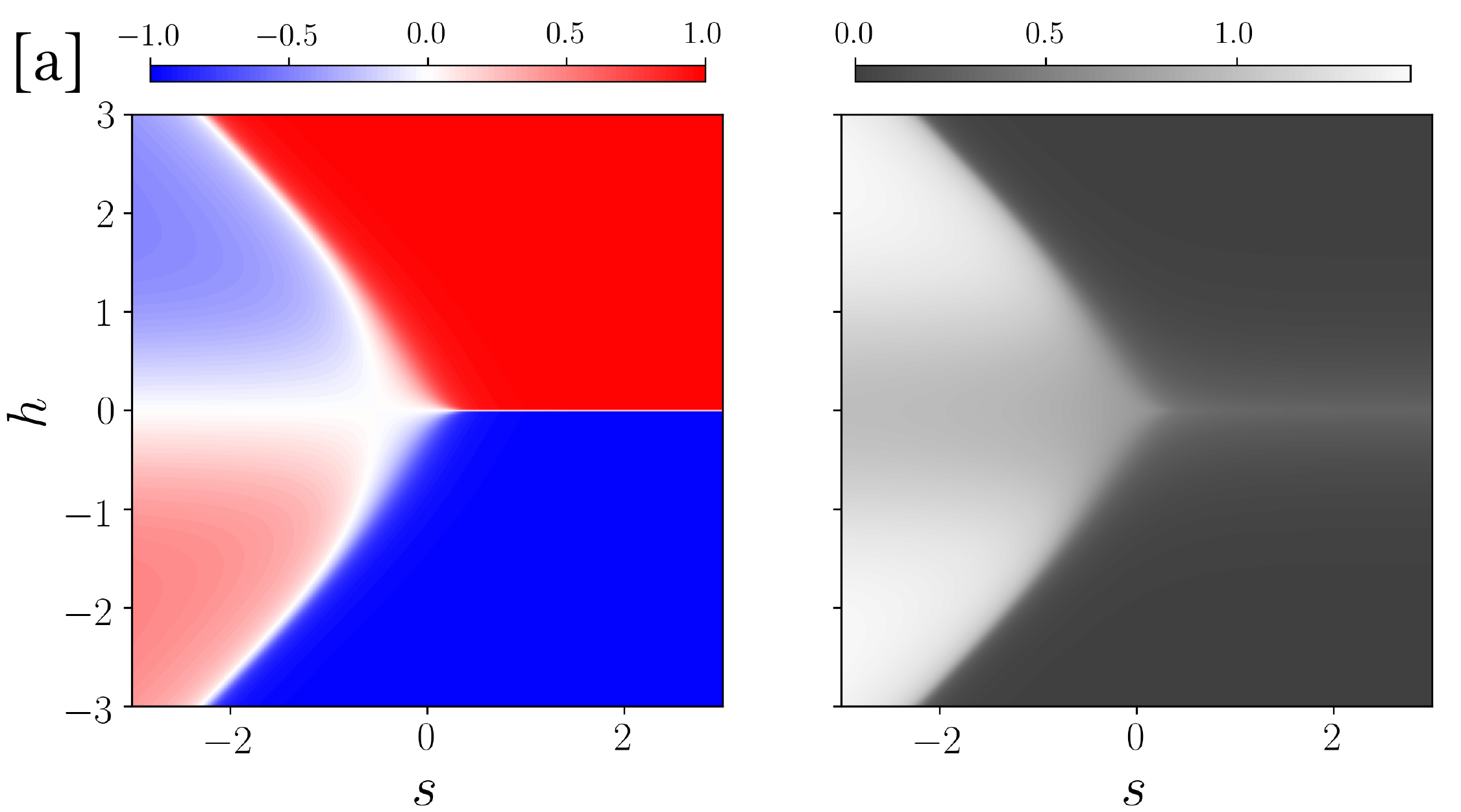}
  \includegraphics[width=0.7\linewidth]{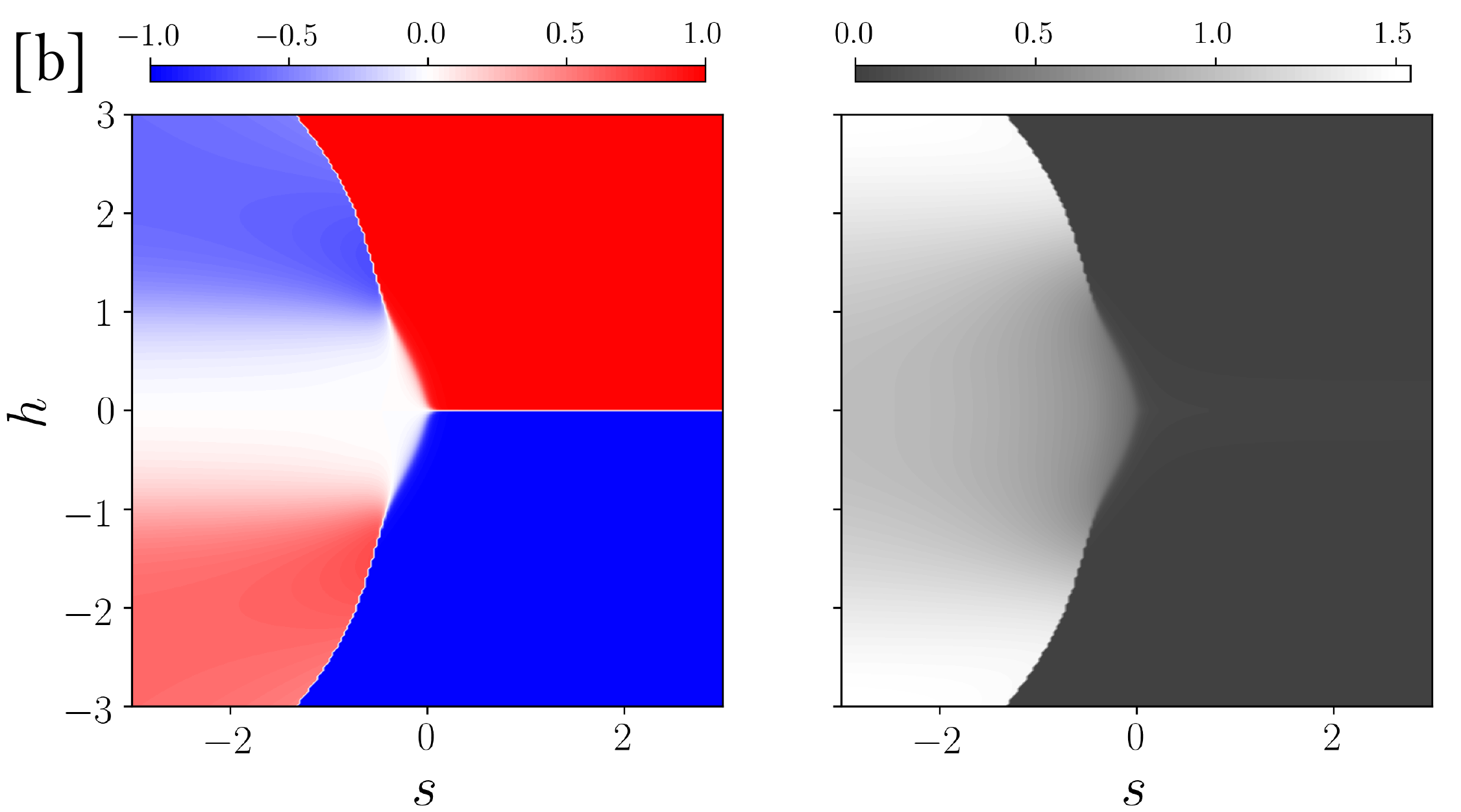}
  \caption{\textbf{ Phase diagrams of the 1D Ising model ($N=12$)}. 
    [\textbf{a}]: $J=0.45$; average magnetisation $\avg{m}{s}$ (\textit{left}),  average escape rate $\avg{r}{s}/N$ (\textit{right}). [\textbf{b}]: $J=1$; average magnetisation $\avg{m}{s}$ (\textit{left}),  average escape rate $\avg{r}{s}/N$ (\textit{right}).}
    \label{fig:phase_diag_ising1d}
\end{figure}

\subsection{Exact diagonalisation}

The exact diagonalisation method is based on the operator ${\cal W}_s$ defined in (\ref{eq:biased_operator}).  This is matrix of size $2^N\times 2^N$.  It can be symmetrised~\cite{jack2010large} by a similarity transform (which leaves its eigenvalues invariant) so it is sufficient to compute the largest eigenvalue of  the symmetric matrix $\widetilde{\mathcal{W}}_{s}$  whose elements are
\begin{equation}
  \label{eq:sym_tilted_operator}
  \left( \widetilde{\mathcal{W}}_{s} \right)_{\bsigma', \bsigma} = e^{\beta E(\bsigma')/2} {\left( \mathcal{W}_{s} \right)}_{\bsigma', \bsigma} e^{-\beta E(\bsigma)/2 } \; .
\end{equation}
One-time observables in the biased ensemble can also be computed from the eigenvector $b$ that corresponds to this largest eigenvalue, in particular the stationary distribution of the biased ensemble is
\begin{equation}
\mu^*(\bsigma) \propto b(\bsigma)^2
\end{equation}
where the constant of proportionality is fixed by normalisation.
We have obtained the eigenvalues and eigenvectors of $\widetilde{\cal W}_s$ up to $N=20$.  
For numerical work, we again set $\beta=1$.

As noted above, the existence of a critical point at $s>0$ and $h=0$ has already been established analytically.    We focus here on the behaviour for $s<0$, to understand if there are critical points in this regime (analogous to those in \fref{fig:phase_diag_mfim}b).
Fig.~\ref{fig:phase_diag_ising1d} summarises the behaviour as a function of $(h,s)$ for two different values of $J$, in a small system ($N=12$).
The situation resembles that of the mean-field case (\fref{fig:phase_diag_mfim}), in particular there is an anomalous regime for $s<0$ where the magnetisation is antiparallel to $h$.  Taking $h>0$ and decreasing $s$ from zero, there is an abrupt crossover from positive to negative $m$, reminiscent of the first-order phase transitions in the mean-field case.

\begin{figure}
  \centering
  \includegraphics[width=0.5\linewidth]{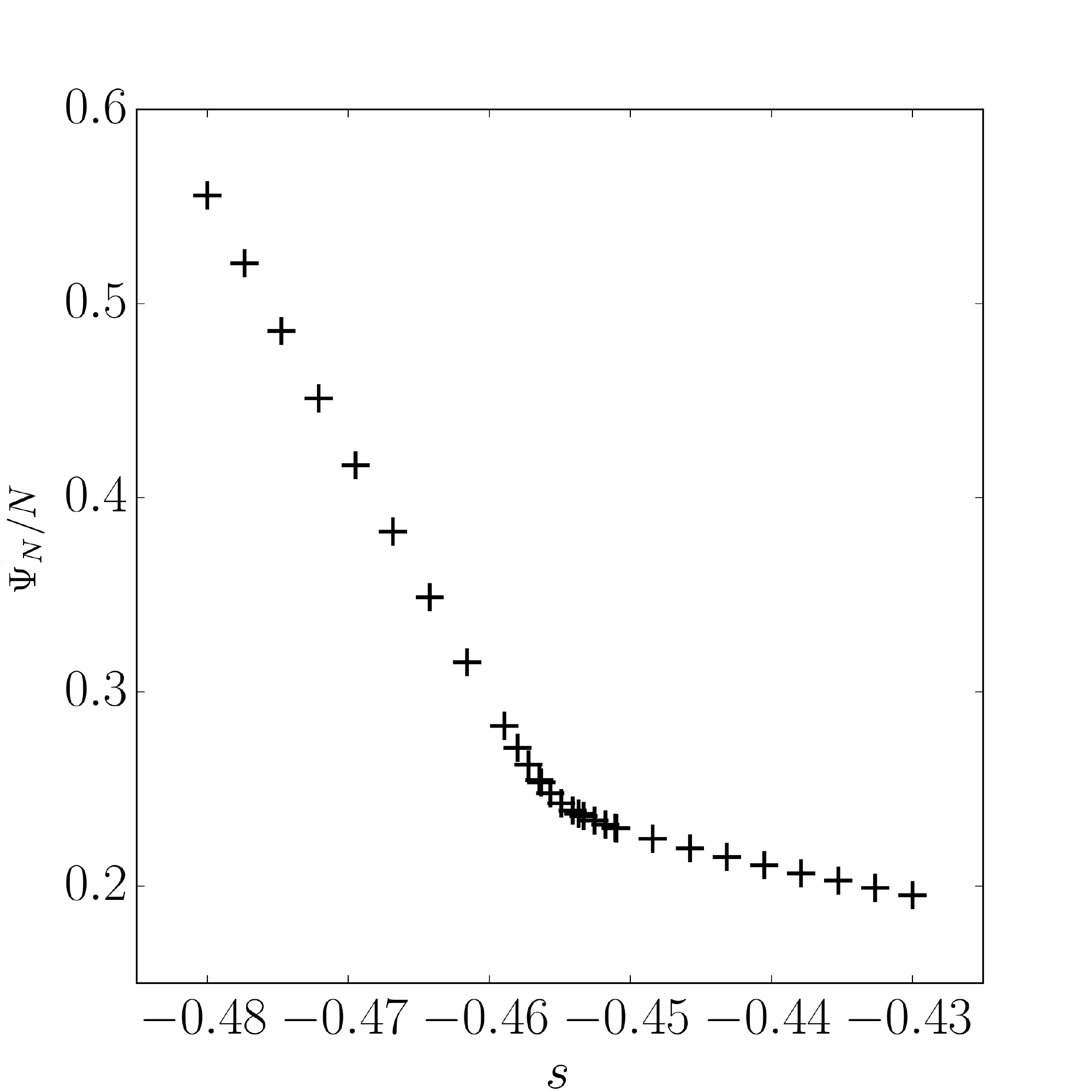}
  \caption{Dynamical free energy $\Psi_{N}(s)/N$ ($N=20$, $J=1$ and $h=1.15$).}
\label{fig:dyn_free_engy}
\end{figure}

\begin{figure}
  \centering
  \includegraphics[width=0.9\linewidth]{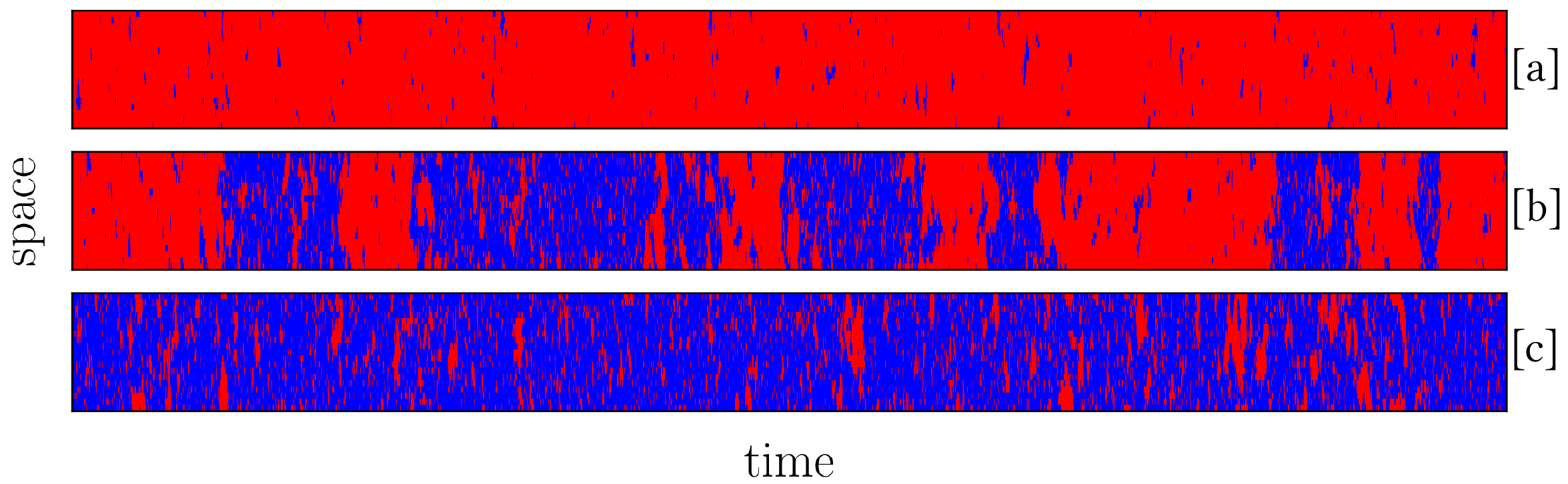}
  \caption{Typical trajectories obtained from Monte-Carlo simulations of the optimally-controlled dynamics ($N=20$, $J=1$, $h=1.15$). \emph{From top to bottom}: [\textbf{a}]  $s=-0.4$ (inactive phase), [\textbf{b}] $s=-0.4557575 \simeq s^{\ast}$ (coexistence), [\textbf{c}] $s=-0.47$ (active phase). The displayed time span is $T=2.1 \times 10^{3}$ ($17068$ Monte-Carlo steps). At coexistence, the trajectory visits both phases, recall \fref{fig:inst2}.}
  \label{fig:config_spacetime}
\end{figure}

\begin{figure}
  \centering
  \includegraphics[width=1.\linewidth]{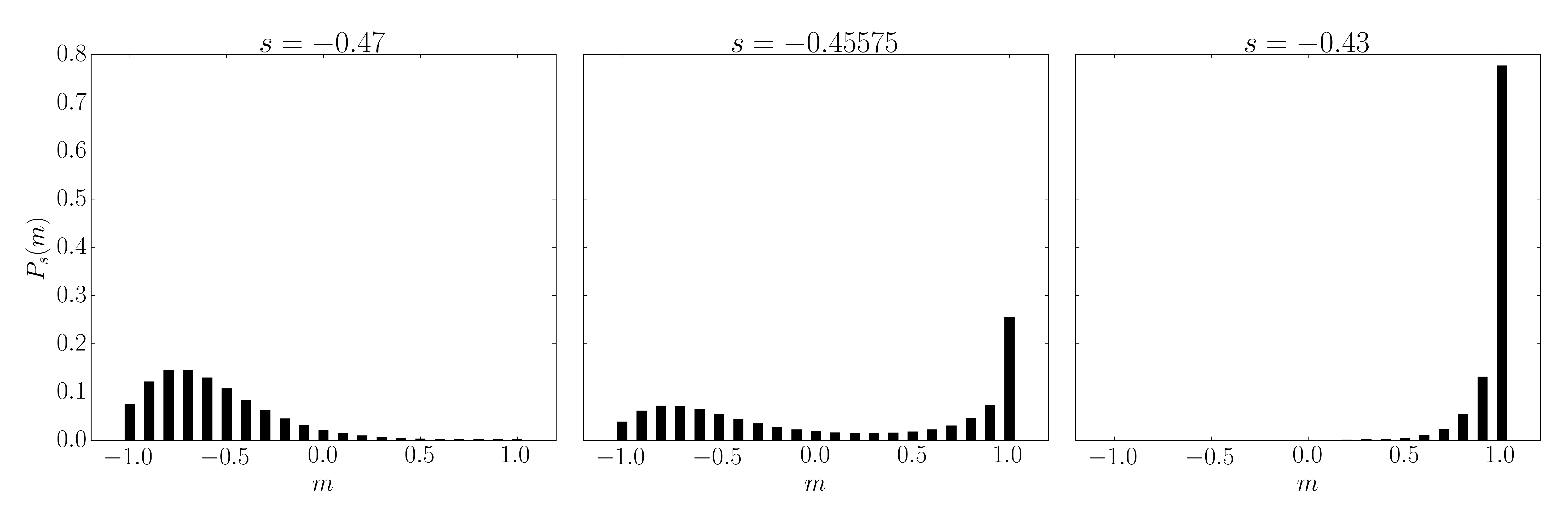}
  \caption{Histograms of the magnetisation $m(t)$ for the 1D Ising model (for $1\ll t\ll T$). Parameters are: $N=20$, $J=1$, $h=1.15$.}
  \label{fig:histo_ising1d}
\end{figure}

\Fref{fig:dyn_free_engy} shows the maximum eigenvalue $\Psi$ as a function of $s$, for $J=1$, $h=1.15$.  
The system is finite so the function $\Psi$ is necessarily analytic, but it does have an abrupt change in slope at $s\approx -0.456$, which is again consistent with the existence of a first-order phase transition.  

The exact diagonalisation also allows exact construction of the optimally-controlled dynamics of \sref{sec:theory_markov_jump}.  This dynamics was simulated by a continuous-time Monte Carlo method, to generate representative trajectories with non-typical values of the activity.  \Fref{fig:config_spacetime} shows examples from the ferromagnetic phase and from the anomalous phase, as well as an example at phase coexistence.  In this last case, the system visits both phases, with rare transitions between them, recall \fref{fig:inst2}.
The anomalous phase also has a non-trivial structure, we return to this point in \sref{sec:anom-1d} below.  
Finally, \fref{fig:histo_ising1d} shows histograms of the magnetisation $m(t)$.  
At the putative point of phase coexistence, this has a bimodal structure, from the two phases.

\begin{figure}
  \centering
  \includegraphics[width=0.55\linewidth]{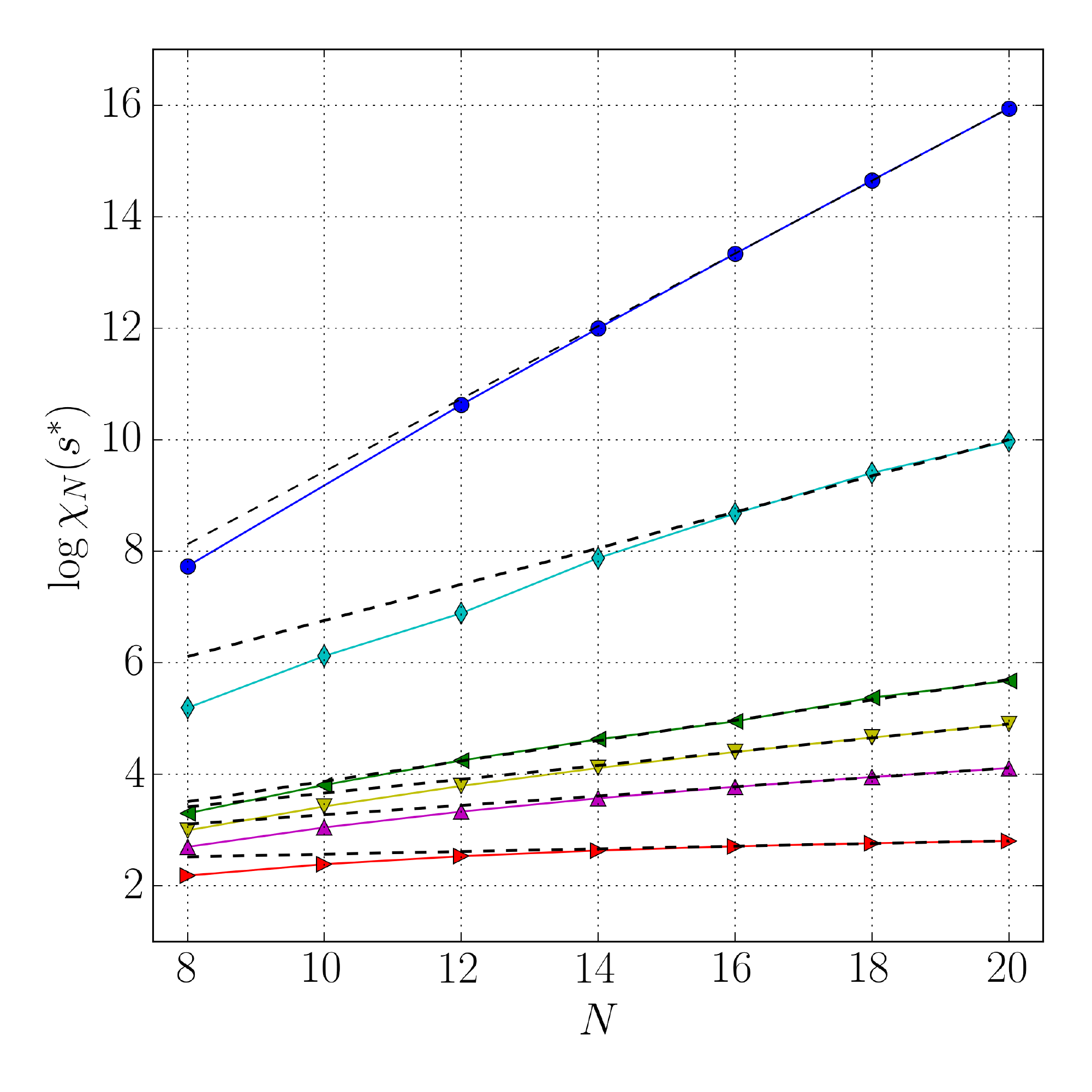}
  \caption{\textbf{$\log\chi_{N}(s^{\ast})$ versus $N$ at $J=1$ for different magnetic fields $h$}. From bottom to top: $h= 1, \, 1.1, \, 1.15, \, 1.2, \, 1.5, \, 2$.}
  \label{fig:chi_vs_L_exact_diago}  
\end{figure}

To establish the existence of a phase transition requires a finite-size scaling analysis.  We concentrate here on the second derivative of the free energy,
\begin{equation}
\chi_N(s) = \Psi''_N(s) \; .
\end{equation}
We recall from (\ref{eq:finite_size_scaling_ddpsi_sstar_mfim}) that $\chi_N(s^*)$ is predicted to diverge exponentially with $N$ in systems at dynamical phase coexistence.  This hypothesis is tested in \fref{fig:chi_vs_L_exact_diago}.
Even at these moderate system sizes an exponential scaling of $\chi_{N}(s^{\ast})$ seems well established for $N\geqslant 14$.  In the next section, we use a cloning method to access larger system sizes and confirm this scaling.
We observe that the derivative of $\log \chi_{N}(s^{\ast})$ with respect to $N$ appears to vanish as $h$ approaches $1$. This indicates that the system is approaching the critical point at the end of the first-order line.  Beyond this point, the first-order transition becomes a smooth crossover.

\subsection{Results --- cloning algorithm with controlled dynamics}

The cloning algorithm introduced in \cite{giardina2006direct, lecomte2007numerical} is a numerical method for sampling trajectories from biased ensembles, and for computing dynamical free energies such as $\Psi_N(s)$.  
We use the implementation described in \cite{brewer2018efficient}, with a fixed population of $N_{c}$ clones, with $N_c$ up to $8\times 10^6$.

The algorithm consists of running the unbiased dynamics independently for each clone, for a time period $\Delta t_{c}$.  This is followed by a cloning step which resamples the population, to account for the biasing factor $e^{ -s \mathcal{A}(t,t+\Delta t_c)}$.

We aim here to sample biased ensembles in relatively large systems, where $s$ is of order unity.  Since the cloning method generates paths according to the original model dynamics, it can be inefficient for sampling ensembles that differ strongly from the model's natural ($s=0$) dynamics.
In order to make the cloning method efficient in such cases, we exploit an alternative formulation of the biased ensemble of trajectories.  We introduce a controlled model with new dynamical rates as in \sref{sec:theory_markov_jump}.   The probability distribution for trajectories of this model is $P^{\mathrm{con}}(\Theta_{\tobs})$ which may be written in the form
\begin{equation}
P^{\mathrm{con}}(\Theta_{\tobs}) = P(\Theta_T) {\rm e}^{-{\cal Q}(\Theta_T)}
\end{equation}
where ${\cal Q}$ is the log-ratio of the trajectory probabilities for the controlled model and the original (Ising) model, for which exact formulae are available~\cite{nemoto2017finite,chetrite2015variational}.  Then $P_s(\Theta_T) = Z(s,T)^{-1} P^{\mathrm{con}}(\Theta_{\tobs}) {\rm e}^{{\cal Q}(\Theta_T)-s{\cal A}(\Theta_T)}$ which has the interpretation of a biased ensemble for the controlled model, which can be sampled by cloning~\cite{nemoto2016population,nemoto2017finite,ray2018exact}.

In the Markov jump framework considered here, the controlled dynamics is defined through its transition rates $w^{\mathrm{con}}(\bsigma' | \bsigma)$. 
\rlj{We take 
\begin{equation}
  \label{eq:controlled_transition_rates}
  w^{\mathrm{con}}(\bsigma' | \bsigma) = a(\bsigma, \bsigma') e^{- (\beta/2) \left[ E^{\mathrm{con}}(\bsigma') - E^{\mathrm{con}}(\bsigma) \right] } \; .
\end{equation}
where $E^{\rm con}$ is the energy of the controlled model, but we emphasise that the mobility $a$ is unchanged from the original model (it depends on $E(\bsigma)$ but not on $E^{\rm con}(\bsigma)$).   This is consistent with \eref{eq:optimal_control_forces_explicit}, we identify $E^{\rm con}=E+U^{\rm con}$.}
For the energy of the controlled model, we take  a functional form with 4 free parameters  $(J_{1}^{\mathrm{c}}, J_{2}^{\mathrm{c}}, K_{3}^{\mathrm{c}}, h^{\mathrm{c}})$, see also~\cite{Jack2014east}:
\begin{equation}
  \label{eq:energy_controlled_system}
  \fl\qquad
  E^{\mathrm{con}}(\bsigma) = - \sum_{i=1}^{N} \left[ J_{1}^{\mathrm{c}} \sigma_{i}\sigma_{i+1} + J_{2}^{\mathrm{c}} \sigma_{i}\sigma_{i+2} + K_{3}^{\mathrm{c}} \sigma_{i}\sigma_{i+1}\sigma_{i+2} + h^{\mathrm{c}} \sigma_{i} \right] \; .
\end{equation}
so that the energy change on flipping spin $i$ depends on its second neighbours in addition to the nearest neighbour interaction of the Ising model.
(We adopt $\sigma_{N+1}= \sigma_{1}$ and $\sigma_{N+2}=\sigma_{2}$ in order to satisfy the periodic boundary conditions.)

\begin{table}
  \centering
  \begin{tabular}{|c|c|c|c|c|}
    \hline
    $s$       &   $J_{1}^{\mathrm{c}}$  & $J_{2}^{\mathrm{c}}$   & $K_{3}^{\mathrm{c}}$  & $h^{\mathrm{c}}$ \\
    \hline
    $-0.44$ (inactive phase)   &   $0.6969$           &  $0.1986$           & $0.0813$           & $0.1998$      \\
    \hline
    $-0.455$ ($\sim$ coexistence)  &   $0.5548$           &  $0.3012$           & $0.1672$           & $-0.1128$      \\
    \hline
    $-0.47$ (active phase)   &   $0.4721$           &  $0.0899$           & $0.0281$           & $-0.2191$       \\
    \hline                                                                         
  \end{tabular}
  \caption{Values of the optimised controlled parameters $(J_{1}^{\mathrm{c}}, J_{2}^{\mathrm{c}}, K_{3}^{\mathrm{c}}, h^{\mathrm{c}})$ for $J=1$, $h=1.15$ and $N=15$.}
  \label{tab:param_controlled_system}
\end{table}

The accuracy of the cloning method requires that the clone population ${\cal N}_c$ is large, otherwise the model suffers from both systematic and random errors~\cite{nemotoFinitetime2017,guevarahidalgoFinitetime2017,rayImportance2018,angeli2019limit}. In principle, the method can yield accurate results whatever the values of the control parameters $(J_{1}^{\mathrm{c}}, J_{2}^{\mathrm{c}}, K_{3}^{\mathrm{c}}, h^{\mathrm{c}})$, but in practice one requires a good choice of these parameters, otherwise the number of clones required for accurate results may be prohibitively large.  Several methods for optimisation of the control parameters have been proposed~\cite{nemoto2016population,nemoto2017finite,ray2018exact,rayImportance2018,nemoto2019optimizing,Dolezal2019}.  Here we use information from exact diagonalisation of small systems to estimate control parameters for larger systems.  

Specifically, recall that $\mu^*(\bsigma)$ is the ($s$-dependent) distribution for configurations in the biased ensemble and let $P^{\rm con}(\bsigma)\propto {\rm e}^{-\beta E^{\rm con}(\bsigma)}$ be the Boltzmann distribution associated with (\ref{eq:energy_controlled_system}).    Also let $\mathrm{KL}(P || Q) $ be the Kullback-Leibler (KL) divergence between two distributions $P,Q$.  Then we maximise the symmetrised KL divergence $\mathrm{KL}(P^{\rm con} || \mu) + \mathrm{KL}(\mu || P^{\rm con})$ over the control parameters and we use these parameters in our controlled model.  The symmetrised KL divergence is small when the two distributions are similar, so this is a practical method for estimation of the control parameters.

\subsubsection{Structure of the  anomalous phase}
\label{sec:anom-1d}
\Tref{tab:param_controlled_system} gives the resulting parameters at several state points.  These reveal useful information about the phases that coexist at $s=s^*$. Even for the paramagnetic phase the $h^{c}$ and $J_{1}^{c}$ are substantially reduced with respect to the natural dynamics, the $J_{2}^{c}$ and $K_{3}^{c}$ are acting to stabilise domains of size $2$ and greater.  As one passes through $s^*$ then $h^{c}$ changes its sign (consistent with the anomalous phase which has a negative response to $h=1.15$).  The couplings promote ferromagnetic order in the system closest to $s^*$, consistent with the fact that there are coexisting phases with opposite magnetisation.

One also observes from \fref{fig:phase_diag_ising1d} that the anomalous phase ($s<0$) experiences a crossover at $h\sim J $ from a magnetisation $m$ close to $0$ to an anomalous (negative) magnetisation when $h$ increases. This can be rationalised by the fact that the mobility \eref{eq:def:glauber_type_transition_rates} is large when the local field $|J(\sigma_{i-1}+\sigma_{i+1}) + h|$ 
experienced by a spin $i$ is the closest to $0$. Its three possible values are $2J+h$, $h$ and $-2J+h$.
\revjg{ For $h$ small ($|h|\ll 2J$), the preferred local field is minimal when $\sigma_{i-1}+\sigma_{i+1}$ is zero. The dominant configuration that}
\rlj{maximises} 
the mobility thus should be $\bsigma=(\dots{}++--++--++--\dots{})$ which has a magnetisation $0$.
\revjg{
 For $h > J$ instead, the local field may not be close to $0$ but local configurations for which $\sigma_{i-1}+\sigma_{i+1} = -2$ display a larger activity than the others. Hence, configurations for which spins are locally surrounded by adjacent spins pointing in the opposite direction than the external field are thus promoted, and the overall magnetisation tends to be negative.}

\rlj{For the mean-field model, we recall that the anomalous phase is close to a \emph{local maximum} of the free energy, when \revjg{$|s|$ and $|h|$ are} small.  Since the argument is based on the fact that $f'_{\rm eq}$ vanishes, one may expect in general large-deviation events may alternatively be localised near saddle points of the free energy.  The character of the anomalous phase in $1d$ is different: this is partly because our numerical results are not able to access anomalous phases at small $|s|$. (To analyse such phases would likely require larger $J$, and in this case larger system sizes would be required.)  In such systems, our expectation is that the anomalous phase would be close to a saddle point of the free energy.  

For example, consider a $1d$ system with many domains that all have equal sizes (as distinct from an equilibrium state where domain sizes are exponentially distributed).  To relax towards equilibrium requires some domain walls to be removed, which happens by diffusion followed by pair-annihilation.  However, the thermodynamic force driving this effect is weak in a system with equally-sized domains, because there are no domain walls that can immediately annihilate, and the motion of individual domain walls is not biased in any particular direction.  Hence, we can interpret this state as a saddle point.  

High-activity states with a sharp distribution of domain sizes have been previously observed in the East model~\cite{Jack2014east}: some similar considerations apply in that case, but domain relaxation times depends very strongly on their sizes in that model, which tend to sharpen the domain-size distribution.  Independent of these details, when comparing the $1d$ case with mean-field, the key message is that identifying saddles or maxima of the free energy is no longer a simple task, but it can still be expected that large deviations with high-activity should be correlated (for small $|s|$) with weak thermodynamic forces.}

\subsubsection{Finite-size scaling}
\label{sec:finite_size_scaling_new}

\begin{figure}
  \centering
  \includegraphics[width=0.49\linewidth]{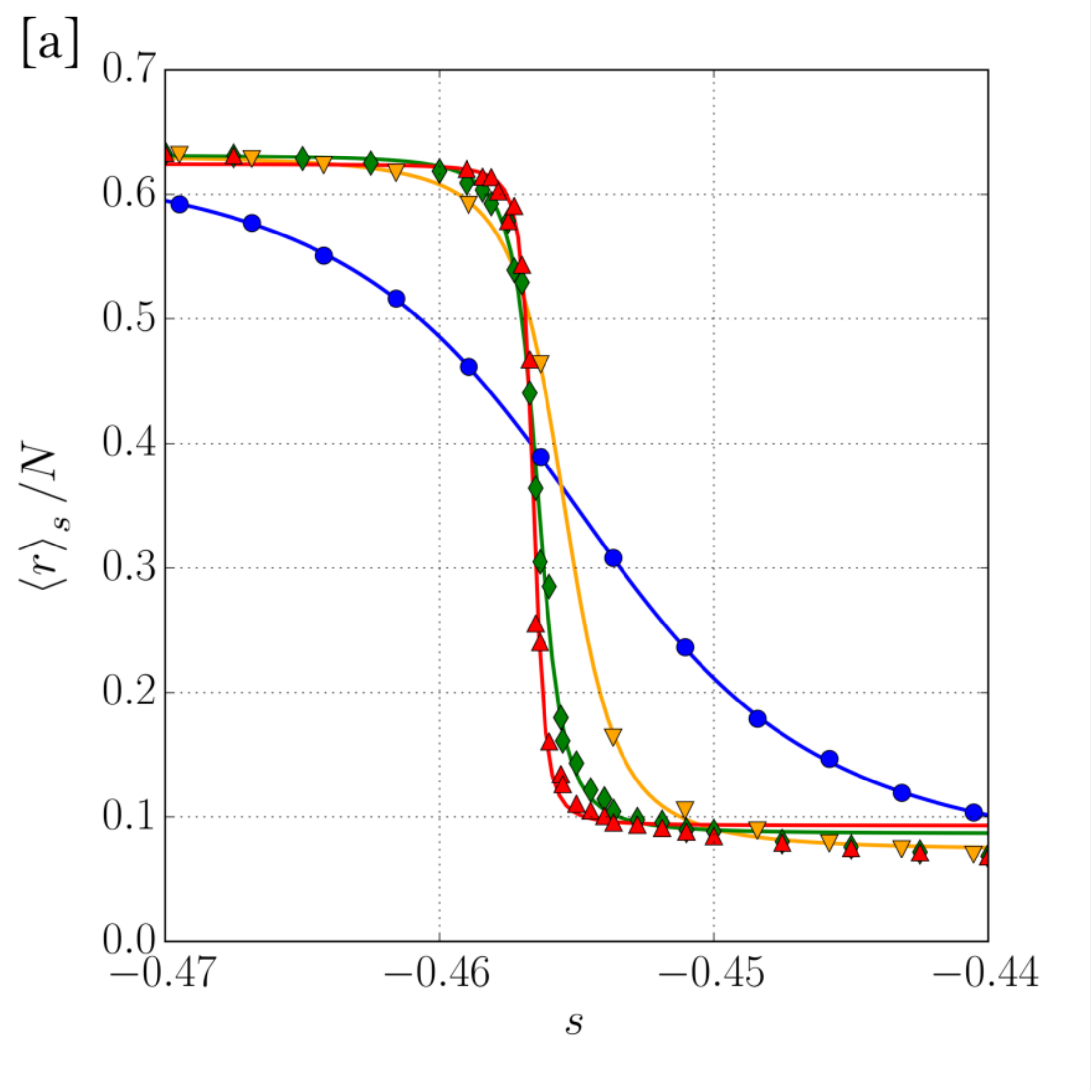}
  \includegraphics[width=0.49\linewidth]{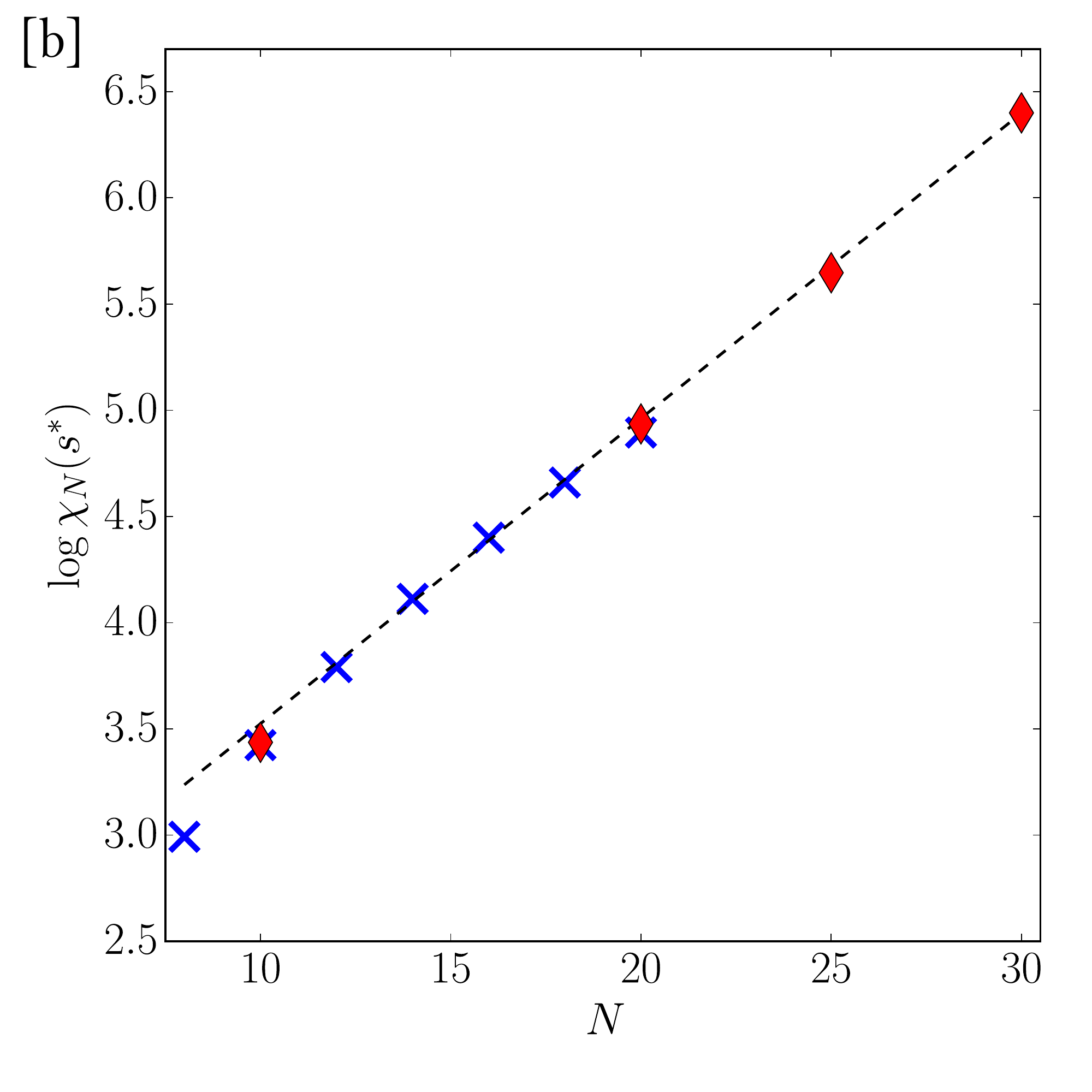}
  \caption{[\textbf{a}]: \textbf{Escape rate per spin vs $s$ from the cloning algorithm}. The magnetic field is $h=1.15$; system sizes are (from the smoothest curve to the steepest one): $N=10, 20, 25, 30$. Continuous lines are the fit obtained from \eref{eq:app:approx_dev_psi} (\ref{app:poisson}).  [\textbf{b}]: \textbf{$\log \chi_{N}(s^{\ast})$ versus $N$ at $J=1$, $h=1.15$}. Blue crosses are obtained from exact diagonalisation. Red diamonds are obtained from the cloning algorithm. The dashed line is a linear fit over $5$ larger $N$ values.}
\label{fig:escape_rate_cloning}
\end{figure}

Using the cloning algorithm with this controlled dynamics, we have obtained results for system sizes up to $N=30$ by using up to $N_{c} \approx 8 \times 10^{6}$ clones and $\tobs \approx 6 \times 10^{3}$. 
Results are shown in \fref{fig:escape_rate_cloning}[a].
We performed five independent runs (with different random seeds) which we use to estimate error bars.  
The error bars are significant close to $s^*$, partly because the fluctuations in the biased ensemble are very large at this point, and also because the optimised control parameters depend strongly on $s$ in this regime, so our method for determining these parameters may not yield the optimal choice for numerical sampling.  The results close to $s^*$ are fitted to the theoretical form \eref{eq:app:approx_dev_psi} which corresponds to the first derivative of (\ref{eq:finite-size-psi}), see \ref{app:poisson} for details.  This allows estimation of the maximal susceptibility $\chi^*$ which is plotted in \fref{fig:escape_rate_cloning}[b].  One sees clear evidence for an exponential growth of this susceptibility, consistent with (\ref{eq:finite_size_scaling_ddpsi_sstar_mfim}) and the arguments of \ref{app:poisson}.

\section{Conclusion}\label{sec:conclusion}

We have considered the activity-biased ensemble of the Ising model evolving with Glauber dynamics. The Mean-Field version of the model has been presented in \sref{sec:mfim}. In addition to what was already known in the absence of magnetic field $h=0$ \cite{lecomte2007thermodynamic}, we find a first order phase transition that occurs for $s<0$ and $h\neq 0$.  The transition separates an inactive ferromagnetic phase ($s>s^{\ast}$) from an anomalous active one ($s<s^{\ast}$). The quasi-potential associated with the stationary probability distribution of the biased ensemble has been characterised analytically.  For the one dimensional Ising model, we have used exact diagonalisation and a cloning algorithm to show that the phase diagram is similar to the mean-field model for $J<J^{\ast}$.  In particular, finite size scaling analysis has confirmed, in accordance with \cite{nemoto2017finite, nemoto2014finite,JackNemoto2019}, an exponential divergence of the correlation function of the time-integrated activity.

\rlj{For general implications of our results, we note that large deviations with low activity tend to be associated with ordered states (ferromagnetic in this case).  In the mean-field model for $J>J^*$, this leads to a symmetry-breaking transition at $s^*>0$, and a similar effect is observed in one dimension.  The mechanisms of large deviations with high activity are more subtle.  In mean-field and for small $|s|$, we find that the system tends to be localised close to a free-energy maximum where the thermodynamic force $f_{\rm eq}'$ is weak.  We have explained that this is likely to be generic in mean-field models, with localisation close to either maxima or saddle points -- one way to see this is that weak control forces are sufficient to localise the system near such states.  In one dimension, it is not so simple to identify the analogs of saddle points, and we are also limited by our numerical methods to phase transitions that occur at relatively large $|s|$.  However, we have explained (in Sec.~\ref{sec:anom-1d}) that localisation of the system in states with weak thermodynamic forces can still be expected in $1d$ settings. 
}

As a final remark on context, we note that these models support Ising-like critical points and associated first-order transitions, with associated phase coexistence.  This situation is widespread in models where large deviations have been studied~\cite{lecomte2007thermodynamic,jack2010large,van2010second,Elmatad2010,Elmatad2013,Turner2015}.  In atomistic models of glasses, the situation is less clear but results are also consistent with first-order transition lines and Ising critical points~\cite{hedges2009dynamic,Pitard2011,Speck2012-sens,Malins2012-sens,Fullerton2013,Turci2017prx}.  The general picture demonstrated here for the one-dimensional Ising model is also expected to apply in those (finite-dimensional) cases -- exponentially diverging susceptibility $\chi(s^*)$ and long trajectories consisting of many domains of each of the coexisting phases~\cite{nemoto2017finite,JackNemoto2019,Jack2019}.  This reinforces the conclusion of~\cite{nemoto2017finite,JackNemoto2019} that this phenomenology is rather general in systems with first-order dynamical phase transitions.  The extent of this generality -- for example its robustness to disorder (as in random-field Ising models and spin glasses~\cite{van2010second,Jack2010rom}) -- is one possible direction for future work.   We note however that while models with slow hydrodynamic modes may exhibit some similar phase transitions~\cite{Tizon2017,baek2017dynamical}, the results of this work are likely not applicable~\cite{Jack2019}.  For example it is notable that Ising models with conserved (Kawasaki) dynamics behave quite differently to models with non-conserved dynamics as considered here: the slow hydrodynamic relaxation of large clusters leads to a diverging time scale and there are singularities in SCGFs already at $s=0$~\cite{loscar2011thermodynamics}.

\ack

We thank Takahiro Nemoto, Jakub Dolezal and Mike Cates for helpful discussions.
This work was funded in part by the Royal Society under grant RP17002.
The figures of this article were made using Matplotlib \cite{Hunter2007}. Numerical analysis were performed using the Python language with the SciPy library \cite{2019arXiv190710121V}. 


\appendix

\section{Computation of $J_{X}$ in the mean-field Ising model for Glauber transition rates}\label{sec:app:computation_J_cross}

Fig.~\ref{fig:phase_diag_mfim}[b] shows a critical point for $s<0$ and $h>0$.  Let the position of this critical point be $(s_{c^+},h_{c^+})$.
This appendix explains that as $J\to J_X$ (from above) then $s_{c^+}\to-\infty$ and $h_{c^+}\to+\infty$, with a fixed value of $(s_{c^+}+h_{c^+})$.
We also derive the value of $J_X$.


To do so, we analyse the minima of the Landau-like free energy $\phi(m,s)$ \eref{eq:phi} in the asymptotic limit $h\to \infty$. For the Glauber transition rates \eref{eq:transition_rates_mfim} [$\gamma(x) = 1/{\cosh}(x)$] we have
\begin{equation}
  \label{eq:app:phi_glauber}
  \phi(m,s) = 1 - m\tanh(2Jm+h) - e^{-s} \frac{\sqrt{1-m^{2}}}{\cosh(2Jm+h)} \; .
\end{equation}
For large $h \gg 1$, $\phi(m,s)$ can be expanded in $\epsilon = e^{-h}$ as
\begin{equation}
  \label{eq:app:expansion_phi}
  \fl
  \eqalign{
    \phi(m,s) \underset{h\gg 2J}{=} (1-m) - \epsilon \, \left[ 2 e^{-s} e^{-2Jm} \sqrt{1-m^{2}} \right] \\
     \qquad \qquad \quad + \epsilon^{2}\left[ 2me^{-4Jm} +  2\epsilon e^{-s}\sqrt{1-m^{2}}e^{-6Jm} \right]
  + O(\epsilon^{3}, e^{-s}\epsilon^{4}) \; .
  }
\end{equation}
Taking $\epsilon\to0$ at fixed $s$ gives $\phi(m,s) \approx 1-m$ in which case $m=1$ is the global minimum (low activity phase).
To obtain the behaviour of $(s_{c^+},h_{c^+})$ we define $\Delta = s+h$ (hence $\epsilon e^{-s} = e^{-\Delta}$) and consider the limit $\epsilon\to0$ at fixed $\Delta$, leading to
%
\begin{equation}
  \label{eq:app:expansion_phi_vs_s}
  \phi(m,s) = (1-m) + e^{-\Delta}c(m)  + \epsilon^{2}k(m,\Delta) + {O}\left(\epsilon^{3} \right) \; .
\end{equation}
with $c(m)=-2e^{-2Jm} \sqrt{1-m^{2}}$ and $k(m,\Delta)=2me^{-4Jm} + 2e^{-\Delta}\sqrt{1-m^{2}}e^{-6Jm}$. 

At leading order then $\phi(m,s) \to (1-m) + e^{-\Delta}c(m)$.  If $c(m)$ is strictly convex then the limiting $\phi$ has a single minimum.  
We observe that
\begin{equation}
  \label{eq:ddc(m)}
  \fl
  c''(m) = \frac{2e^{-\Delta}e^{-2Jm}}{(1-m^{2})^{3/2}} \big[ -4J^2m^4 + 8J^2m^2 + 4Jm^3 - 4J^2 - 4Jm + 1 \big] \, .
\end{equation}
Analysing this expression numerically indicates that it is indeed strictly convex for small $J$.  However for $J=J_X\approx 0.402964$ we have $c''(m^*)=0$ (for some point $m^*$), and $c''(m)>0$ elsewhere.  In addition, taking $e^\Delta=c'(m^*)$ and $\epsilon\to0$ in (\ref{eq:app:expansion_phi_vs_s}) ensures that $\partial \phi / \partial m=0$  at $m^*$.  That is, there is a stationary point of $\phi$ with vanishing curvature.  This is a critical point with (formally) $h_{c+}=\infty$ and $s_{c^+}+h_{c^+}=\Delta$, as asserted above.

For $J<J_X$ the convexity of $c$ means that $\partial ^2\phi / \partial m^2>0$ and there are no phase transitions at these large values of $h$.  For $J>J_X$ then $c$ is not convex and it follows that $\phi$ has two (local) minima (as $\epsilon\to0$, for $\Delta$ in some suitable range).  Moreover, there is a choice of $\Delta$ for which these minima are degenerate, which corresponds to dynamical phase coexistence as $h\to\infty$ (with fixed $\Delta$).  For finite $h$ (that is, $\epsilon\neq0$) the line of first-order coexistence can be traced in the $(s,h)$ plane until it ends at a critical point at $(s_{c^+},h_{c^+})$. 

This establishes the picture advertised above: the crossover between the situations shown in Figs.~\ref{fig:phase_diag_mfim}[a,b] occurs by the critical points for $s<0$ (in panel [b]) moving away from $(s,h)=(0,0)$; as $J\to J_X$ (from above) then they diverge as $(s_{c^+},h_{c^+})\to(-\infty,+\infty)$.  Hence they are absent in panel [a].

\section{Finite size scaling and Poisson process approximation}\label{app:poisson}

{For completeness, this appendix details the Poisson process approximation for dynamical phase coexistence already introduced in \cite{nemoto2017finite,JackNemoto2019}.}

As mentioned at the end of section~\ref{sec:quasipot_instanton} and in section~\ref{sec:finite_size_scaling_new}, the effective dynamics at the coexistence point $s=s^{\ast}$ can be approximated as a bi-stable process of parameter (transition rate) $\omega_{0}$.

Hence, one can think that the behaviour of the biased ensemble for $s$ very close to $s^{\ast}$ can be well approximated by a new biased ensemble obtained from the Poisson approximation valid at $s=s^{\ast}$. For this purpose, one introduces a two states Markov jump process $x(t)\in \{1, 2\}$.




According to the two states approximation, one can write that
\begin{equation}
  \label{eq:app:two_states_approx_psi}
  \Psi_{N}(s) \underset{s \sim s^{\ast}}{\approx} \frac{1}{N} \lim_{\tobs \to \infty} \frac{1}{\tobs} \avg{ e^{-(s-s^{\ast})N\int_{0}^{\tobs}a(x(t))\dst t}}{\mathrm{Poiss}}
\end{equation}
where $\avg{}{\mathrm{Poiss}}$ is referring to the average with respect to the two states approximation of the effective dynamics at $s=s^{\ast}$ and $a(x)\in\{a_{1},a_{2}\}$ is the activity of the system.

In order to compute the right hand side of \eref{eq:app:two_states_approx_psi}, one can easily solve the tilted eigenvalue problem for this simple bi-stable system. One can indeed easily see that
\begin{equation}
  \label{eq:app:transfert_matrix_formulation}
   \avg{e^{-(s-s^{\ast}) \int_{0}^{\tobs}a(x(t)) \dst t }}{\mathrm{Poiss}} = \sum_{x',\, x} {\left(e^{\tobs \mathcal{W}_{s}^{\ast}}\right)}_{x',\, x} \, p_{0}(x) \; ,
\end{equation}
with $p_{0}$ the initial distribution and $\mathcal{W}_{s}^{\ast}$ the tilted matrix that reads as
\begin{equation}
  \label{eq:app:def:transfert_matrix}
  \mathcal{W}_{s}^{\ast}  = \left(
\begin{array}{cc}
  -(s-s^{\ast})Na_{1} - \omega_{0} & \omega_{0} \\
  \omega_{0} & -(s-s^{\ast})Na_{2} - \omega_{0}
\end{array}
\right) \;~.
\end{equation}

In the large time limit $\tobs\to\infty$, the right hand side of \eref{eq:app:transfert_matrix_formulation} is dominated by $\exp(N\tobs\lambda_{\max}^{\ast}(s))$ where $N\lambda_{\max}^{\ast}(s)$ is referring to the largest eigenvalue of $\mathcal{W}_{s}^{\ast}$. The eigenvalue per unit system size $N$ thus reads
\begin{equation}
  \label{eq:app:largest_eigenval_transfert_matrix}
  \fl
  \lambda_{\max}(s)  = -(s-s^{\ast})  \frac{a_{1} + a_{2}}{2} - N^{-1}\omega_{0} + \sqrt{(s-s^{\ast})^{2}\frac{(a_{2} - a_{1})^{2}}{4} + N^{-2}\omega_{0}^{2}}
  \; .
\end{equation}
Hence, $\Psi_{N}(s) \approx \lambda_{\max}^{\ast}(s)$ for $s$ close to $s^{\ast}$ and one obtains in particular
\begin{equation}
  \label{eq:app:approx_dev_psi}
  \fl
  \Psi_{N}'(s) \underset{s\sim s^{\ast}}{\approx} \frac{ \dst \lambda_{\max}^{\ast}}{\dst s}(s) = -\frac{a_{1}+a_{2}}{2} + \frac{1}{2} \frac{(s-s^{\ast}){(a_{2} - a_{1})}^{2}}{\sqrt{(s-s^{\ast})^{2}(a_{2}-a_{1})^{2} + 4N^{-2}\omega_{0}^{2}}} \; ,
\end{equation}
and
\begin{equation}
  \label{eq:app:approx_ddev_psi_sstar}
  \Psi_{N}''(s^{\ast}) \approx \frac{ \dst^{2} \lambda_{\max}^{\ast}}{\dst s^{2}}(0) = N \frac{(a_{2}-a_{1})^{2}}{4}\omega_{0}^{-1} \ \, .
\end{equation}

Equations (\ref{eq:app:approx_dev_psi}, \ref{eq:app:approx_ddev_psi_sstar}) thus proves (\ref{eq:finite-size-psi}, \ref{eq:finite_size_scaling_ddpsi_sstar_mfim}).





\section{Exact solution for the 1D Ising model in the absence of external magnetic field ($h=0$)}\label{app:exact_ising1D}

In the absence of any magnetic field $h$ ($h=0$), one can diagonalize the tilted operator $\mathcal{W}_{s}$ by mapping the latter into the Hamiltonian of a quantum Ising chain \cite{jack2010large},  see also~\cite{garrahan2009first}. The derivation follows closely those of \cite{jack2010large,garrahan2009first} we also correct two small errors present in~\cite{jack2010large}.

We first transform from spin variables $\sigma$ to domain wall variables.
In one dimension, the presence of a domain wall between site $i$ and $i+1$ can be quantified by defining $\tau_{i} = \case{1}{2}(1-\sigma_{i}\sigma_{i+1})$ such that $\tau_{i}=1$ if there is a domain wall ($\sigma_{i}\sigma_{i+1}=-1$) and $\tau_{i}=0$ if not ($\sigma_{i}\sigma_{i+1}=1$). We have $h=0$ so the energy $E = -J\sum_{i=1}^{N} \sigma_{i}\sigma_{i+1}$ reads as $E=J\sum_{i=1}^{N}\left( 2\tau_{i} - 1 \right)$.

In terms of the domain wall variables $\tau_{i}$, flipping one spin (say $\sigma_i$), transforms the configuration $\btau = {\{ \tau_{i} \}}_{i=1}^{N}$ into $\btau' = \{\tau_{1}, \dots{}, 1-\tau_{i-1}, 1-\tau_{i}, \tau_{i+1}, \dots{}, \tau_{N} \}$. We denote the associated transition rate by $w(\btau'|\btau)$. 
For Glauber transition rates then
\begin{equation}
  \label{eq:app:glauber_dw_var}
    w(\btau'|\btau) = 1 + \nu \left[ \left( \tau_{i}-\case{1}{2} \right) + \left( \tau_{i-1}-\case{1}{2} \right) \right] \; ,
\end{equation}
where we have introduced $\nu = \tanh (2\beta J)$.

The second step is to represent the symmetric tilted operator $\widetilde{\mathcal{W}}_{s}$ \eref{eq:sym_tilted_operator} in terms of elementary operators. Since the spin variable on each site is a two state variable $\tau_{i}=0,1$, we work with Pauli matrices: Let
\begin{equation}
  \label{eq:app:def:pauli_matrices}
  S_{i}^{+} = \left(
    \begin{array}{cc}
      0 & 1 \\
      0 & 0 
    \end{array}
  \right) \; ,
  \quad
  S_{i}^{-} = \left(
    \begin{array}{cc}
      0 & 0 \\
      1 & 0 
    \end{array}
  \right) \;~,
\end{equation}
 which are linked to the Pauli matrices $S_{i}^{x}$, $S_{i}^{y}$, $S_{i}^{z}$ through the usual relations $S_{i}^{\pm} = \case{1}{2}\left(S_{i}^{x} \pm i S_{i}^{y} \right)$ and $S_{i}^{z}=2S_{i}^{+}S_{i}^{-}-1$. Note in particular $S_{i}^{+}S_{i}^{-} \left| \tau_{i} \right\rangle = \tau_{i} \left| \tau_{i} \right\rangle$. The tilted operator \eref{eq:sym_tilted_operator} thus reads
\begin{equation}
  \label{eq:app:tilted_op_pauli_mat}
  \eqalign{
    \widetilde{\mathcal{W}}_{s} = \sum_{i=1}^{N}\, & e^{-s}\left[ S_{i-1}^{+}S_{i}^{-} + S_{i-1}^{-}S_{i}^{+} + \sqrt{1-\nu^{2}}\left( S_{i-1}^{+}S_{i}^{+} + S_{i-1}^{-}S_{i}^{-}\right) \right] \\
    &  - \nu \left[ S_{i-1}^{+}S_{i-1}^{-} + S_{i}^{+}S_{i}^{-} \right] + \nu - 1
    } \; ,
\end{equation}
with the periodic boundary conditions $S_{0}^{\pm}= S_{N}^{\pm}$.

The third step uses a Jordan-Wigner transformation to map the spin operators $S_{i}^{\pm}$ into fermionic creation and annihilation operators $f_{i}^{\dag}$ and $f_{i}$ respectively. Hence we arrive at a quadratic Hamiltonian (free fermions) which is easy to diagonalize. For a system of size $N=1$, both spin and fermionic operators would be the same. However, for $N>1$, one must introduce the Jordan-Wigner transformation \cite{sachdev_2011,grynbergExact1994}: for $i\geq2$ we take
\begin{equation}
  \label{eq:app:jordan_wigner_transfo}
  \eqalign{
    f_{i}^{\dag} = e^{-i\pi\sum_{k=1}^{i-1}S_{k}^{+}S_{k}^{-}} S_{i}^{+}\\
     f_{i} = e^{i\pi\sum_{k=1}^{i-1}S_{k}^{+}S_{k}^{-}} S_{i}^{-}  \; .
    }
\end{equation}
Also $ f_{1}^{\dag} = S_{1}^{+}$ and $f_{1} = S_{1}^{-} $.  These $f_i$ satisfy the standard fermionic anticommutation relations.
Also, $f_{i}^{\dag}f_{i} = S_{i}^{+}S_{i}^{-}$. 

Now, for $i\geq2$ we have $f_{i-1}^{\dag}f_{i} = S_{i-1}^{+}S_{i}^{-} \exp(i\pi S_{i-1}^{+}S_{i-1}^{-}) = S_{i-1}^{+}S_{i}^{-}$ [the second equality can be verified by Taylor expansion of the exponential and using that $(S_{i-1}^{+})^2=0$].  Similarly
 $S_{i-1}^{-}S_{i}^{+} = - f_{i-1}f_{i}^{\dag}$ and $S_{i-1}^{+}S_{i}^{+}=f_{i-1}^{\dag}f_{i}^{\dag}$ and $S_{i-1}^{-}S_{i}^{-}=-f_{i-1}^{}f_{i}^{}$.
 However, the terms that involve hopping across periodic boundaries require additional care.
 Write
\begin{equation}
 f^\dag_N f_1 = e^{-i\pi ({\cal N}_{\rm dw} - S_N^+ S_N^-)} S_{N}^{+}  S_1^- = - e^{-i\pi {\cal N}_{\rm dw}} S_{N}^{+}  S_1^- 
\end{equation}
where ${\cal N}_{\rm dw} = \sum_{k=1}^{N} S_{k}^{+}S_{k}^{-}$ is the operator for the total number of domain walls.
[We used that ${\cal N}_{\rm dw}$ commutes with $S_N^+ S_N^-$ and that $e^{-i\pi S_N^+ S_N^-} S_{N}^{+} = - S_{N}^{+}$.]
Now observe that since we consider an Ising model with periodic boundaries, the number of domain walls in the system is always an even number.  
Hence all state vectors $|\psi\rangle$ in the space where $\widetilde{\mathcal{W}}_{s}$ operates have $e^{-i\pi {\cal N}_{\rm dw}}|\psi\rangle = |\psi\rangle$, so the term
$S_{N}^{+}  S_1^-$ in $\widetilde{\mathcal{W}}_{s}$ can be identified with $ - f^\dag_N f_1$ after the Jordan-Wigner transformation. 

It will be convenient to introduce anti-periodic boundary conditions for the fermion operators, so we introduce the notation
\begin{equation}
  \label{eq:boundaries_rel_fermionic_op}
  f_{0}^{\dag} = -f_{N}^{\dag} \; , \qquad f_{0} = - f_{N}  \; .
\end{equation}
Within $\widetilde{\mathcal{W}}_{s}$ we may then identify $S_{N}^{+}  S_1^-=  f^\dag_0 f_1$ and (similarly)
$S_{0}^{-}S_{1}^{+} = -f_{0}f_{1}^{\dag}$ and $S_{0}^{+}S_{1}^{+} = f_{0}^{\dag}f_{1}^{\dag}$ and $S_{0}^{-}S_{1}^{-} = -f_{0}f_{1}$.
Collecting terms we find
\begin{equation}
  \label{eq:app:fermionic_tilted_op}
  \eqalign{
    \widetilde{\mathcal{W}}_{s} = \sum_{i=1}^{N} & \left\{ e^{-s}\left[f_{i-1}^{\dag}f_{i} - f_{i-1}f_{i}^{\dag} + \sqrt{1-\nu^{2}} \left(f_{i-1}^{\dag}f_{i}^{\dag} -f_{i-1}f_{i}\right) \right] \right. \\
      & \;\: \left. - 2\nu f_{i}^{\dag}f_{i} + \nu - 1 \right\}
    } \; .
\end{equation}

The next step of the derivation is to diagonalise this quadratic operator, which is done in  two stages. 
%
One first makes a Fourier transform 
\begin{equation}
  \label{eq:fourier_transform}
  \hat{f}_{q} = \frac{1}{\sqrt{N}} \sum_{j=1}^{N} f_{j} e^{iqj} \; , \qquad f_{j} = \frac{1}{\sqrt{N}} \sum_{q\in Q_{N}^{\rm odd}} \!\! \hat{f}_{q} e^{-iqj} \; ,
\end{equation}
where $Q_{N}^{\rm odd} = \left\{ q = \case{\pi}{N}(2k+1) : k \in \{ -\frac{N}{2}, 1-\frac{N}{2},\dots,\frac{N}{2}-1\} \right\}$.
Odd wavenumbers have been chosen in order to satisfy the anti-periodic property of $f_{j}$ and $f_{j}^{\dag}$ \eref{eq:boundaries_rel_fermionic_op} (itself a consequence of the conserved parity of the number of domain walls which is always even). One can easily show that $\hat{f}_{q}$ and $\hat{f}_{q}^{\dag}$ are fermionic operators (which satisfy fermionic anti-commutation rules). The symmetrised tilted operator \eref{eq:app:fermionic_tilted_op} then reads
\begin{equation}\label{eq:app:fourier_transform_fermionic_tilted_op}
  \fl
  \eqalign{
    \widetilde{\mathcal{W}}_{s}  =  \sum_{q\in Q_{N}^{\rm odd}} & \left\{ e^{-s} \left[ e^{-iq} \hat{f}_{q}^{\dag}\hat{f}_{q} - e^{iq} \hat{f}_{q}\hat{f}_{q}^{\dag}  + \sqrt{1-\nu^{2}}\left(e^{-iq} \hat{f}_{q}^{\dag}\hat{f}_{-q}^{\dag} - e^{iq} \hat{f}_{q}\hat{f}_{-q} \right) \right] \right. \\
    &\:\: \left. -2\nu \hat{f}_{q}^{\dag} \hat{f}_{q} + \nu - 1 \right\} \; ,
    }
\end{equation}
which can be recast as
\begin{equation}
  \label{eq:app:fourier_transform_fermionic_tilted_op_simplified}
  \fl
  \widetilde{\mathcal{W}}_{s} = \!\!\! \sum_{q\in Q_{N}^{\rm opp}} \!\! \left\{
    2\left(e^{-s}\cos(q) - \nu \right) \hat{f}_{q}^{\dag}\hat{f}_{q} - i\sqrt{1-\nu^{2}}e^{-s}\sin(q)
    \left(\hat{f}_{q}^{\dag}\hat{f}_{-q}^{\dag} - \hat{f}_{-q}\hat{f}_{q} \right) + \nu - 1
    \right\}
\end{equation}
by using $\hat{f}_{q}\hat{f}_{q'} = -\hat{f}_{q'}\hat{f}_{q}$ and $\hat{f}_{q}\hat{f}_{q}^{\dag} + \hat{f}_{q}^{\dag}\hat{f}_{q} = 1$ and $\sum_{q} \cos(q) \hat{f}_{-q}\hat{f}_{q} = \sum_{q} \cos(q) \hat{f}_{q}^{\dag}\hat{f}_{-q}^{\dag}= 0$, according to the fermionic commutation rules.

Finally, we follow the standard procedure for diagonalising the Hamiltonian of a 1D quantum-Ising chain in a magnetic field (see for instance \cite[section 10.1]{sachdev_2011}) in which this same operator appears: define
\begin{equation}
  \label{eq:app:linear_transfo_fq}
  \hat{c}_{q} = \cos(\theta_{q}) \hat{f}_{q} - i\sin(\theta_{q})\hat{f}_{-q}^{\dag}\; ,
\end{equation}
with $\theta_{-q}=-\theta_{q}$. Taking $\theta_{q}$ such that $\tan(2\theta_{q}) = \sqrt{1-\nu^{2}}\sin(q)/(\cos(q)-\nu)$, we find
\begin{equation}
  \label{eq:app:sym_tilted_op_fermionic_diag}
  \widetilde{\mathcal{W}}_{s} = \! \sum_{ q \in Q_{N}^{\rm odd} } \!\! \left[ \Lambda_{s}(q) - 1\right ] - \!\! \sum_{q\in Q_{N}^{\rm odd}} \!\! 2\Lambda_{s}(q) \hat{c}_{q}\hat{c}_{q}^{\dag} \; ,
\end{equation}
with $\Lambda_{s}(q)=\sqrt{(e^{-s}\cos(q) - \nu)^{2} + (1-\nu^{2})e^{-2s}\sin(q)^{2}}$.
The maximum eigenvalue of \eref{eq:app:sym_tilted_op_fermionic_diag} is thus associated with the full occupied state $\otimes_{q} \left| 1 \right\rangle$ and reads
\begin{equation}
  \label{eq:largest_eigenvalue_fermionic}
  \Psi_N(s) = \sum_{q\in Q_{N}^{\rm odd}} \!\! \left( \Lambda_{s}(q) - 1 \right) \; .
\end{equation}
One verifies that $\Psi(0) = 0$ as it must be.

One can also compute the critical value $s_{c}$ for which $\psi(s)=\lim_{N\to\infty}\Psi_N(s)$ displays a second order singularity. By studying the sum \eref{eq:largest_eigenvalue_fermionic} that tends toward a continuous integral in the $N\to\infty$ limit, it has been shown \cite{jack2010large} that the singularity happens at $s_{c}$ for which $\lim_{q\to 0}\Lambda_{s_{c}}(q) = 0$. This yields\begin{equation}
  \label{eq:app:s_star_exact_h0}
  s_{c} = - \ln \nu = -\ln \tanh(2 \beta J) \; .
\end{equation}

The results of this section are very similar to those of  \cite[Sec 4.4.2]{garrahan2009first}: we have clarified the use of odd wavenumbers when summing $q$ and we have noted the existence of a transition at $s=s_c$.
[Note $\Lambda$ may be rearranged as $\sqrt{(e^{-2s}-1)(1-\nu^2) + (1-\nu e^{-s}\cos q)^{2}}$.]
Comparing to \cite[Eq. (3.2)]{jack2010large}, the calculation is slightly different because they considered large deviations of the energy and not the activity.  The use of odd wavenumbers was not discussed in that~\cite{jack2010large}.  Following the method described here as outlined in  Appendix B of~\cite{Jack2019} recovers their results, which corrects a factor of 2 in~\cite{jack2010large}.



\newpage
\section*{References}

\bibliographystyle{unsrt}
\bibliography{refs_art_activity_ising_vfin}

\end{document}